\documentclass{article}
\usepackage[cp1250]{inputenc}
\usepackage[T1]{fontenc}
\usepackage{times}
\usepackage{enumerate}
\usepackage{amsmath}
\usepackage{amsthm}
\usepackage{amsfonts}
\usepackage{graphicx}
\usepackage{textcomp}
\usepackage{lscape}
\usepackage{a4wide}
\usepackage{array}
\usepackage{color}
\usepackage[normalem]{ulem}
\newtheorem{lem}{Lemma}
\newtheorem{rem}{Remark}
\newtheorem{cor}{Corollary}
\newtheorem{thm}{Theorem}

\newtheorem{example}{Example}
\newtheorem{prop}{Proposition}

\newcounter{athm}
\newenvironment{athm}{\ \\  \noindent \addtocounter{athm}{1} \textbf{Theorem \Alph{athm}.}  \it }{ \ \ }
\newcounter{acor}

\newcommand{\mat}{\theta}

\newcommand{\Hil}{U} 
\newcommand{\normh}[1]{\abs{#1}_\Hil} 
\newcommand{\binh}[1]{{\Big\< #1 \Big\>_\Hil}} 
\newcommand{\inh}[1]{{\left\< #1 \right\>_\Hil}} 

\newcommand{\set}[1]{{\{#1\}}}
\newcommand{\be}{\begin{equation}}
\newcommand{\ee}{\end{equation}}
\newcommand{\beq}{\begin{eqnarray*}}
\newcommand{\eeq}{\end{eqnarray*}}
\newcommand{\beqa}{\begin{eqnarray}}
\newcommand{\eeqa}{\end{eqnarray}}
\newcommand{\wsp}{ \ \ \ \ \ \ } 
\def\cadlag{c\`adl\`ag }
\def\prt{{\partial }}
\def\defi{\smash{\mathop{=}\limits^{\Delta}}}
\def\abs#1{|#1|}
\def\f#1#2{\frac{#1}{#2}}
\def\la{\lambda}
\def\a{\alpha }
\def\d{\delta }
\def\s{\sigma }
\def\th{\theta }
\def\Om{\Omega }
\def\De{\Delta }
\def\R{{\mathbb  R}}
\def\P{{\mathbf  P}}
\def\F{{\cal F}}
\def\EE{{\bf  E}}
\def\B{\mathcal{B}}
\def\H{{\cal H}}
\def\G{{\cal G}}
\def\La{\Lambda}
\def\I{1\!\!1}
\def\>{\rangle}
\def\<{\langle}
\def\ol#1{\overline{#1}}
\def\td#1{\tilde{#1}}
\def\h{\hat }
\def\n{\noindent }
\def\sqr#1#2{{\vcenter{\vbox{\hrule height.#2pt
       \hbox{\vrule width.#2pt height#1pt \kern#1pt
       \vrule width.#2pt} \hrule height.#2pt}}}}
\def\kd{\par\nobreak{\hfill$\sqr83$}\par\medbreak}

\title{
Defaultable bonds with an infinite number of L\'evy factors
\thanks{Research supported in part by Polish KBN Grant P03A 034 29
``Stochastic evolution equations driven by L\'evy noise''.} }
\author{Jacek Jakubowski,
\thanks {Institute of Mathematics, University
of Warsaw, ul. Banacha 2, 02-097 Warszawa, Poland and Faculty of
Mathematics and Information Science,  Warsaw University of
Technology {\tt jakub@mimuw.edu.pl }.} \and
Mariusz Niew\k{e}g\l owski,
\thanks { Faculty of Mathematics and
Information Science, Warsaw University of Technology,   Plac
Politechniki 1, 00-661 Warszawa, Poland {\tt
M.Nieweglowski@mini.pw.edu.pl}.} . } 
\begin{document}
\date{}

\maketitle
  \thispagestyle{empty}

\noindent

\noindent{\bf Abstract } A market with defaultable bonds where the
bond dynamics is in a Heath-Jarrow-Morton setting and the forward
rates are driven by an infinite number of  L\'evy factors  is
considered. The setting includes rating migrations driven by a
Markov chain. All basic types of recovery are investigated. We
formulate necessary and sufficient conditions (generalized HJM
conditions) under which the market  is
arbitrage free. Connections with consistency conditions are
discussed.

\noindent{\bf Keywords } L\'evy processes,  defaultable bonds, HJM
postulate, credit risk, rating migration, conditional Markov
chains

\

\noindent{\bf Mathematics Subject Classification (2000) }
 60H30 \ 91B28 \ 60G51

\

\noindent{\bf JEL Classification } E43 \  G12

\section*{Introduction}

\indent The paper is concerned with the market containing a
risk-free bond and  defaultable bonds issued by companies. A
defaultable bond will default with a certain probability before or
at maturity time $T$.
 The probabilities of defaults depend on
economic conditions of the firms and are reflected by {\it rating
classes} designated by rating agencies like Standard$\&$Poor's or
Moody's. If a default does not occur an owner of the bond receives,
as in the case of default-free bond, one currency unit. In the
case of default the owner obtains a part of the promised payoff. This
part depends on the credit rating of the issuer of the bond 
 and on the
adopted recovery scheme.
  To model   defaultable bonds we use the intensity based
  models which are the basic way of modeling (see e.g. Bielecki and Rutkowski
\cite{Bielecki2000}, Lando \cite{Lando04}).
  In contrast to most papers on the subject, which use
Brownian motion for modeling (see e.g. Duffie and Singleton
\cite{Duffie1999}, Bielecki and Rutkowski \cite{Bielecki2000},
\cite{Bielecki2004}), we apply the theory of L\'evy processes which
admit discontinuous trajectories and contain many standard
processes like Brownian motion, Poisson processes, and generalized
hyperbolic L\'evy motion.

 It is well known that using   L\'evy processes to
modeling has many advantages (see e.g. Eberlein and \"{O}zkan
\cite{Ozkan2003}, Eberlein and Kluge \cite{Eberlein2007}, Cont and
Tankov \cite{Cont2007}, \"{O}zkan and Schmidt \cite{Ozkan04}) such as
better calibration procedure for  real world and also risk
neutral data. Eberlein and  Raible \cite{Eberlein1999} and
Eberlein and \"{O}zkan \cite{Ozkan2003} used  finite dimensional
L\'evy processes with exponential moments in some
neighborhood of zero  to model the term structure of defaultable
forward rates. They generalize the  approach of Bielecki and
Rutkowski \cite{Bielecki2000} to defaultable bonds with rating
migration. This approach is in the spirit of Heath, Jarrow and
Morton (hereafter HJM) methodology \cite{HJM}. They assume that
 real-world defaultable forward rates dynamics as well as
recovery schemes are exogenously specified and they establish
existence of an arbitrage free model that supports
these objects. More precisely, they show that if the intensity
matrix process satisfies the so called "consistency condition"
then one can construct a rating migration process and price
processes  of defaultable bonds with credit migration that are,
under an appropriate measure,  local martingales after discounting.
The consistency conditions are interpreted as  conditions on the
intensity matrix of the rating migration process. We should stress
that these conditions do not determine  the intensity matrix
proces uniquelys, so actually there can be infinitely many transition
matrix processes satisfying those systems of equations. Neither Bielecki
and Rutkowski nor Eberlein and \"Ozkan  attempt to
generalize the HJM condition to a condition on the drift
term which guarantees that the HJM postulate is satisfied, i.e.
that the discounted bond prices are local martingales.  In this paper we
do  this  in the case of defaultable bonds.  We
cover all situations of practical importance.
 The same question in the infinite dimensional case was considered
by Schmidt \cite{Schmidt05} with Brownian motion as a noise and by
\"{O}zkan and Schmidt \cite{Ozkan04} with L\'evy noise and  recovery
of market value.
 \cite{Schmidt05}  gives necessary
and sufficient conditions for discounted prices of defaultable
bonds to be martingales in the case of rating based recovery of
market value and   recovery of treasury value.
 \"{O}zkan and Schmidt's \cite{Ozkan04}
approach is based on Musiela parameterization and requires more
stringent conditions on the model than ours, since the It\^o
formula for processes with values in Hilbert spaces is used. As we
notice in Remark \ref{uw3j} their result is not true without some
additional assumptions.

In this paper we give the generalized  HJM conditions  in the case
of defaultable bonds and typical recovery schemes.
 We consider fractional recovery of market value, fractional
recovery of treasury value and fractional recovery of par value.
The multiple default case introduced by Schonbucher
\cite{Schonbucher1998}, which allows one to consider company
reorganization, is discussed as well. From the very beginning  we
assume that the L\'evy processes may be infinite dimensional. The
importance of treating models with an infinite number of factors
was stressed in recent papers of Carmona and Tehranchi
\cite{Carmona2004}, Ekeland and Taflin \cite{Taflin2005}, Cont
\cite{Cont2005} and Schmidt \cite{Schmidt05}.

In Section 1 we recall
basic facts on forward rates driven by L\'evy  processes and the
HJM-type condition for non-defaultable bonds provided that the
market is arbitrage-free. Next, in Section 2, we describe credit
risk models with and without rating migration. The rating classes
vary according to a conditional continuous time Markov chain and
the default time is equal to the time of entering  the worst
rating class. In the main part (Section \ref{sec-HJM}) we give
HJM-type conditions for defaultable bonds with credit migration.
These conditions depend on the form of recovery and the rating
migration process. From a structural point of view, all equations
follow a similar pattern, where one has the classical HJM drift
condition plus an additional term, depending on the particular
recovery payment.
 All is proved under a natural assumption on the default
risk-adjusted short-term interest rate (hypothesis (H1)). More
precisely, under hypothesis (H1) we prove that in the general case
the HJM postulate is equivalent to the HJM condition. It is worth
 mentioning that in a model in which all processes are continuous,
we do not need  to assume (H1). Namely, (H1) plus the HJM condition is
equivalent to the HJM postulate.
 We also formulate HJM conditions in terms of the derivative of the
logarithm of the moment generating function of the L\'evy noise
(Section 4). These forms are much more useful in applications (see
e.g. \cite{JakubNiewe2007}).
 In Section \ref{sect5} we formulate,
following \cite{Bielecki2002}, consistency conditions involving the
recovery structure, default intensities and bond prices. We prove
that these conditions  are equivalent to the HJM type conditions
derived in the previous sections. Hence, under (H1), we can extend
and  generalize to the case of infinite dimensional L\'evy
processes the results of \cite{Bielecki2000} and \cite{Ozkan2003}.
The proofs of our results are given in the last section of the paper.
 The present paper is a revised and significantly extended
version of the preprint \cite{JakubNieweZabcz2005}.

 Summing up, the main contributions of the paper are
necessary and sufficient conditions (generalized HJM conditions)
for the coefficients in the definition of the forward rates ensuring
that the discounted prices of defaultable bonds are martingales.
 These conditions are given for all typical recovery schemes and
with infinite dimensional L\'evy noise as the source of uncertainty
in the dynamics of defaultable forward rates, which is the most
general L\'evy noise one can use. Our assumptions on the  L\'evy
processes are weaker than  having
exponential moments in some neighborhood of zero, as  in
\cite{Ozkan2003}, \cite{Eberlein2007} and \cite{Ozkan04}.

\section{Preliminaries}
We will consider processes on a complete probability space $(\Om, \F
,\P)$. We take L\'evy processes with values in some abstract
separable Hilbert space $\Hil$ as the source of uncertainty in a
model. Let  $Z$ be  a L\'evy process, i.e. a c\` adl\` ag process
with independent and stationary increments and values in $\Hil$ with
inner product denoted by  $\inh{ \cdot, \cdot }$. Let $\F_{t}^{0} =
\sigma(Z(s); s\leq t)$ be the $\sigma$-field generated by $Z(t),
t\geq 0$, and $\F_{t}$ be the completion of $\F_{t}^{0}$ by all sets
of $\P$ probability zero. It is known that this filtration is right
continuous, so it satisfies the "usual conditions".
 We can associate with $Z(t)$ a
measure of its jumps, denoted by $\mu$, i.e. for any $A \in
\B(\Hil)$ such that $\ol{A} \subset \Hil \backslash  \{ 0 \} $ we
have
\[
    \mu([0,t], A) = \sum_{ 0 < s \leq t} \I_{A} (\De Z(s)).
\]
The measure $\nu$ defined by
\[
    \nu (A) = \EE ( \mu([0,1], A)  )
\]
is called the L\'evy measure of the process $Z$. Stationarity of
increments implies that $
    \EE ( \mu([0,t], A)  ) = t \nu (A).
$ The L\'evy-Khintchine formula shows that the characteristic
function of  the L\'evy process has the form
\[
    \EE e^{i \inh{ \la , Z(t) } } = e^{t \psi(\la)},
\]
with
\[
    \psi(\la) = i \inh{ a, \la } - \f{1}{2} \inh{ Q \la, \la }
              + \int_\Hil (e^{i \inh{\la, x}} -1 -i \inh{\la,x} \I_{\{\normh{x} \leq 1 \} } (x) ) \nu(dx),
\]
where $a \in \Hil$, $Q$ is a symmetric non-negative nuclear operator
on $\Hil$, $\nu$ is a measure on $\Hil$ with $\nu( \{ 0 \} ) = 0 $
and
\[
    \int_\Hil (\normh{x}^2 \wedge 1) \nu(dx) < \infty.
\]
Let $b$ be the Laplace transform of $\nu$ restricted to the
complement of the ball $\{y: \,|y|\leq 1\}$,
\begin{equation}     \label{eq:def-b}
b(u) = \int_{{\normh{y}}>1} e^{-{\inh{u, y}}} \nu (dy),
  \end{equation}
and set
$$
B = \{u\in U: \,\, b(u) < \infty\}.
$$
$Z$ has a well known L\'evy-It\^o decomposition:
\[
    Z(t) = a t + W(t)
    + Z_0(t),
\]
where
\[
    Z_0(t) =
    \int_0^t \int_{ \normh{ y} \leq 1} y (\mu(ds, dy) - ds \nu (dy))
    +
    \int_0^t \int_{ \normh{ y} > 1} y \mu(ds,dy),
\]
 and $W$ is a Wiener process with values in $\Hil$ and
covariance operator $Q$.

    Let
    $r(t)$, $t\geq 0$, be the  short rate process and
    $$
    B_t = e^{\int_0^t r(\sigma) d\sigma}.
    $$
    Let $B(t,\theta)$, $0 \leq t \leq \theta \leq T^*$, be the market price at
time $t$ of a risk-free bond paying $1$ at maturity time $\theta$;
$T^*$ is a finite horizon of the model. The
    forward rate curve is a function $f(t,\theta)$ defined for $\ t
    \leq \theta$ and such that
    \begin{equation}
        \label{eq:forward-rate}
        B(t,\theta) = e^{-\int_t^{\theta} f(t, s) ds}.
    \end{equation}
    It is convenient to assume that once a bond has matured its
    cash
    equivalent goes to the bank account.  Thus $B(t,\theta)$, the
    market price at time $t$ of a bond paying $1$ at maturity
    time $\theta$,  is also defined for $t \geq \theta$ by the formula
    \begin{equation}
        \label{eq:bond-price}
        B(t,\theta) = e^{\int_{\theta}^t r(\sigma) d\sigma}.
    \end{equation}
We postulate here the following dynamics for forward rates:
\be\label{eq:HJM-forward-rates-dyn}
    d f(t,\mat) = \a ( t, \mat) dt + \inh{ \s(t,\mat), d Z(t) },
\ee where for each $\theta$ the processes $\alpha(t,\theta)$,
$\sigma(t,\theta)$, $t\leq \theta$, are assumed to be predictable
with respect to a given filtration $({\cal F}_t)$ and such that the
integrals in \eqref{eq:HJM-forward-rates-dyn} are well defined.
 Sometimes we use another form of SDE for forward rates,
\be\label{eq:HJM-forward-rates-dyn-l2}
    d f(t) = \td{\a}(t) dt + \td{\s} (t)  d Z(t),
\ee where $ \td{\a}(t)$ is a function on $[0,T^*]$ given by $
\td{\a}(t)(\mat) = \a (t,\mat) $ and $\td{\s}(t)$ is a linear
operator from $\Hil$ into $L^2[0,T^*]$ defined by
\[
    (\td{\s} (t) u )(\mat) =  \inh{ \s(t, \mat) ,u } .
\]  For $t > \mat$ we put
 \be\label{eq:umowa}
    \a(t,\mat) = \s(t,\mat) = 0 .
\ee
 So we will assume that for given $T^*$, the integrals in
the definition of $f$ exist in the sense of the Hilbert space $H =
L^2[0,T^*]$ with scalar product $( \cdot, \cdot )$.
 We will
regard the coefficients $\alpha$ and $\sigma$ as, respectively, $H$-
and $L(U, H)$-valued predictable processes.

 It follows from \eqref{eq:HJM-forward-rates-dyn} that for $t \leq
 \mat$,
\[
    f(t,\mat ) =  f(0,\mat) + \int_{0}^t \a(s,\mat) ds
    + \int_0^t \inh{\s (s,\mat) , d Z(s) },
\]
and for $t \geq \mat$, according to (\ref{eq:umowa}),
\[
    f(t,\mat ) =  f(0,\mat) + \int_{0}^\mat \a(s,\mat) ds
    + \int_0^\mat \inh{\s (s,\mat) , d Z(s) } .
\]
Thus the process $f(t,\mat)$ for $t \geq \mat$ is constant for
each $\mat > 0$, say equal to $f(\mat,\mat)$, and it can be
identified with the short rate process
\[
    r(\mat) =
    f(\mat,\mat) =
    f(0,\mat) + \int_{0}^\mat \a(s,\mat) ds
    + \int_0^\mat \inh{\s (s,\mat) , d Z(s) }.
\]
\emph{The HJM postulate} states that the discounted bond prices
\[
    \h{B}(t,\mat) = \f{B(t,\mat)}{ B_t }
\]
are local martingales for each $\mat \in [0,T^*]$. Since for $t
> u$ we have $ f(t,u) = f(u,u)$, it follows that
\[
    B_t = e^{\int_{0}^t f( u,u) du} =  e^{\int_{0}^t f( t,u) du},
\]
and thus the discounted bond prices can be written as
\[
    \h{B}(t,\mat)
    = e^{ - \int_{t}^\mat f(t,u) du} e^{ - \int_{0}^t f(t,u) du}
    = e^{ - \int_{0}^\mat f(t,u) du},
\]
and hence the HJM postulate is that the processes
$\h{B}(\cdot,\mat)$ , $\mat \in [0,T^*]$, given by
\[
    \h{B}(t,\mat) = e^{ - \int_{0}^\mat f(t,u) du}
    = e^{ - \left( f(t) , \I_{[0, \mat ]} \right)},
\]
are local martingales. We will assume that the processes $Z$, $\a$
and $\s$ satisfy the following conditions :

\n \textbf{A1a: } The processes $\alpha$ and $\sigma$ are
predictable and  with probability one have  bounded trajectories
(the bound may depend on $\omega$).

\n \textbf{A1b: } For arbitrary $ r>0$ the function $b$ given by
\eqref{eq:def-b} is bounded on  $ \{u: |u| \leq r, b(u)<\infty
\}$.

\n \textbf{A2: }
 For all $\theta \leq T^*,\,$ ${ \P}$- almost surely,
\begin{equation}\label{3K12}
\int_t^\theta \sigma(t,v)\,dv \in B
\end{equation}
 for almost all $ t\in [0, \theta]$.

  It is convenient to express the HJM condition
in terms of the Laplace exponent of the L\'evy process $Z$, i.e. of
the logarithm of the moment generating function of the  process
$Z$, that is, the functional $J:
  \Hil \rightarrow \R $ given by
\beq
    J(u) = - \inh{ u,a } + \f{1}{2} \inh{ Q u ,u }  &+& \int_\set{\normh{y } \leq 1}
    \big(
     e^{ -   \inh{ u,y } } - 1 + \inh{u,y}
    \big) \nu(dy) \\
         &+ & \int_\set{\normh{y } > 1}  \big( e^{ - \inh{ u,y } } - 1 \big)
         \nu(dy).
\eeq
 The following theorem, under other assumptions,
goes back to the paper \cite{Bjork1997} by Bj\"{o}rk, Di Massi,
Kabanov and Runggaldier (see also Eberlein and \"{O}zkan
\cite{Ozkan2003}). We present it following Jakubowski and Zabczyk
\cite{JakubowskiZabczyk2005}.
\begin{athm}
    Assume (A1) and (A2). The discounted bond prices are local martingales if and only if the
following HJM-type condition
    holds:
\be\label{eq:HJM-type-condition-J}
 \int_{t}^{\mat} \a (t,v)  dv = J \bigg( \int_{t}^{\mat} \s(t,v) dv
 \bigg),
\ee for each $\mat \in [0, T^*]$ and almost all $t \ \leq \mat$.
\end{athm}

\n Using integration by parts and the dynamics of the discounted
bond, we obtain
\begin{thm}
    The processes of discounted price of the bond  have the following
    dynamics:
    \begin{align*}
         d\hat{B} &(t,\mat) =
        \hat{B}(t-,\mat)
                \bigg(
        \bar{a}(t,\mat)
        dt \\
        & + \,
       \int_\Hil
       \big[
       e^{
       -
       \inh{ \int_t^\mat \s(t,v) dv, y
       }} -1
       \big] (\mu(dt, dy) - dt \nu (dy))
              -
        \inh{
       \int_t^\mat \s(t,v) dv
        ,
        dW(t)
        }
        \bigg),
    \end{align*}
  where
    \beq
        \bar{a}(t,\mat)
        &=&
        -
        ( \I_{[0,\mat]} ,\td{\a}(t)  )
        +
        J (\int_t^\mat \s(t,v) dv ) .
        \eeq
\end{thm}
\begin{cor}
The process of discounted bond price can be written in the following
integral form:
 \beq
    \hat{B}(t, \mat ) = \hat{B}(0,\mat)
    \exp  \bigg(
    - \int_{0}^{t} ( \I_{[0,\mat]} ,\td{\a}(s) ) ds
    - \int_{0}^{t} \inh{ \int_s^\mat \s(s,v) dv , d Z(s)}
    \bigg),
\eeq and if the HJM-type condition (\ref{eq:HJM-type-condition-J})
holds, then \beq
    \hat{B}(t, \mat ) = \hat{B}(0,\mat)
    \exp \bigg(
    - \int_{0}^{t} J\Big( \int_s^\mat \s(s,v) dv  \Big) ds
        - \int_{0}^{t} \inh{ \int_s^\mat \s(s,v) dv , d Z(s)}
    \bigg).
\eeq
\end{cor}

In what follows we assume that condition
(\ref{eq:HJM-type-condition-J}) is fulfilled.

\section{Description of credit risk models}

In the default-free world, by a bond maturing at time $\mat$ with
face value $1$ we mean a financial instrument whose payoff is $1$ at
time $\mat$. In a defaultable case we have several variants
describing the amount and timing  of so called \emph{recovery
payment} which is paid to bond holders if default has occurred
before the bond's maturity. If we denote by $\tau$ the default time,
then, generally speaking, the payoff of the defaultable bond is as
follows:
\[
    D(\mat,\mat) = \I_{\{ \tau > \mat \}} + \I_{\{\tau \leq  \mat \} } \cdot \text{ \emph{recovery payment}
    }.
\]
If  $\d$ is a  recovery rate process, then a \emph{recovery
payment} can take different forms (see e.g. \cite{Bielecki2002}
and references there):
\begin{itemize}
\item $ \d(t) D( \tau-,\mat) \f{B_\mat}{B_\tau}  $  -- \emph{fractional recovery of market value} -- at time of default
bondholders receive a fraction of the pre-default market value of
the defaultable bond (i.e. of $D(\tau - ,\mat)$):
  \[
    D(\mat,\mat) = \I_{\{ \tau > \mat \}} + \I_{\{\tau \leq  \mat \} } \cdot \d_\tau D( \tau-,\mat)
    \f{B_\mat}{B_\tau} ,
  \]
  where $\d(t)$ is an $\mathbb{F}$- predictable process with values in $[0,1]$.
  \item $ \d $ -- \emph{fractional recovery of Treasury value} --  a fixed fraction $\d$
  of the bond's face value is paid to bondholders at the bond's maturity date $\mat$:
  \[
    D^\d(\mat,\mat) = \I_{\{ \tau > \mat \}} + \I_{\{\tau \leq  \mat \} } \cdot
    \d .
  \]
  \item $ \f{\d B_\mat}{B_\tau}  $ -- \emph{fractional recovery of par
  value}--
 a  fixed fraction $\d$ of the bond's face value is paid to
  bondholders at default time $\tau$:
  \[
    D^\De(\mat,\mat) = \I_{\{ \tau > \mat \}} + \I_{\{\tau \leq  \mat \} } \cdot \d  \f{B_\mat
    }{B_\tau}.
  \]
\end{itemize}

Our objective is to derive the HJM drift condition for models with
different kinds of recovery and with  migration of credit ratings.

\subsection{Models with rating migration}

We give a short description of a model with rating
migration; for details see Bielecki, Rutkowski
\cite{Bielecki2002}. We assume that the credit rating migration
process $C^1$, which is a \cadlag process,  is modeled by a
conditional Markov chain relative to  $\mathbb{F}$ with values
in the set of rating classes $\mathcal{K} = \{1, \ldots, K \}$,
where state $i=1$ represents the highest rank,  $i= K-1$ the
lowest rank, and $i=K$ the default event.
 With state $i$, $i \leq K-1$, there is associated the pre-default term
structure $g_i$. We assume that the instantaneous  defaultable forward rates
have dynamics $g_i(t, \mat)$ given by
\[
    d g_i( t, \mat) = \a_i(t,\mat) d t +\inh{ \s_i(t,\theta), d Z_i(t)
    },  \quad i \in \{1, \ldots, K-1 \},
\]
where $Z_i(t)$ are L\'evy processes with values in $\Hil$. By the
L\'evy-It\^o decomposition,  each $Z_i(t)$ has the form
\[
    Z_i(t) = a_i t + W_i(t)
    +
    \int_0^t \int_{ \normh{ y} \leq 1} y (\mu_i(ds, dy) - ds \nu_i (dy))
    +
    \int_0^t \int_{ \normh{ y} > 1} y \mu_i(ds,dy),
\]
where $a_i \in \Hil$ and $\mu_i$ is the jump measure of $Z_i$. Let
$D_i(t,\mat) = e^{- \int_t^\mat g_i(t,u),du }$ and denote the
discounted values of $D_i$
 by
$\h{D}_i(t,\mat) = \f{D_i (t, \mat)}{B_t}$. Applying the It\^o lemma
 as in the default free case we have (below
$J_i$ corresponds to $Z_i$ in the same way as $J$ corresponds to
$Z$).
\begin{thm}\label{thm:dynamic-pre-default-TS}
The dynamics of the process $\hat{D}_i(t,\mat)$ is given by
    \beq
        d \hat{D}_i(t,\mat) && =
        \hat{D}_i(t-,\mat)
                 \bigg(
            \big(
        g_i
        (t,t) - f(t,t)
        +
        \bar{a}_i (t,\mat)
        \big)
         dt \\
        && +
       \int_\Hil
       \left[
       e^{
       -
       \inh{
       \int_t^\mat \s_i(t,v) dv ,
         y
       }} -1
       \right] (\mu_i(dt, dy) - dt \nu_i (dy))
         -
        \inh{
       \int_t^\mat \s_i(t,v) dv
        ,
        dW_i(t)
        }
        \bigg) ,
    \eeq
    where $\bar{a}_i(t,\mat)$ satisfies
    \begin{eqnarray}\label{eq:drift-ai}
        \bar{a}_i(t,\mat)
        =
        -
        (  \I_{[0,\mat]} ,\td \a_i(t))
        + J_i \left(\int_t^\mat \s_i(t,v) dv \right) .
    \end{eqnarray}
\end{thm}

To preserve the interpretation of rating classes, i.e. the fact that
higher rated bonds are more expensive than lower rated ones, it is
reasonable to assume that
\[
    g_{K-1} (t,\mat) >  g_{K-2} (t,\mat) > \ldots > g_{1} (t,\mat) > f(t, \mat)
\]
for all $t \in [0, \mat ]$ and all $\mat \in [0, T^*]$.
This condition implies that inter-rating spreads
are positive.

If there are given two filtrations $\mathbb{F}$ and $\mathbb{G}$,
then the $\mathbb{F}$-conditional infinitesimal generator of the
process $C^1$ describing the credit rating migration at time $t$
given the $\sigma$-field $\F_t$  has the form
\[
    \La (t) =
    \begin{pmatrix}
      \la_{1,1}(t)   & \la_{1,2}(t)   & \cdots   & \la_{1,K-1}(t)   & \la_{1,K}(t) \\
      \la_{2,1}(t)   & \la_{2,2}(t)   & \cdots   & \la_{2,K-1}(t)   & \la_{2,K}(t) \\
      \vdots         & \vdots         & \ddots   & \vdots           & \vdots   \\
      \la_{K-1,1}(t) & \la_{K-1,2}(t) & \cdots   & \la_{K-1,K-1}(t) & \la_{K-1,K}(t) \\
      0              & 0              & \cdots   & 0                & 0        \
    \end{pmatrix}
\]
where  the off-diagonal processes $\la_{i,j}(t)$, $i \neq j$, are
non-negative processes adapted to $\mathbb{F} \subseteq \mathbb{G}$,
and the diagonal elements are negative and determined by
off-diagonals
\[
    \la_{i,i}(t) = - \sum_{j \in \mathcal{K} \backslash \{i\} }
    \la_{i,j}(t) .
\]

For our purposes we specify $\mathbb{G} = \mathbb{F} \vee
\mathbb{H}$ where $\mathbb{F} = \mathbb{F}^{\hat{Z}} \vee
\mathbb{F}^\Lambda$, $\hat{Z}=(Z, Z_1, \ldots, Z_K)$ and $\mathbb{H}
= \mathbb{F}^{C^1} $, i.e. $\F_t = \F^{\hat{Z}}_t \vee
\F_t^{\Lambda} $, $\ \G_t = \F^{\hat{Z}}_t \vee \F_t^{\Lambda} \vee
\F_t^{C^1}$ . A detailed construction of $C^1$ in this case can be
found in Bielecki and Rutkowski \cite{Bielecki2002},
\cite{Bielecki2004} or Lando \cite{Lando1998}.

To describe the credit risk we also need, besides the credit
migration process $C^1$ defined above, the process $C^2$ of the
previous ratings. If we denote by $\tau_1, \tau_2, \tau_3, \ldots $
the consecutive jump times of the credit migration process $C^1$,
then for $ \ t \in [ \tau_{k} , \tau_{k+1} )$,
 \beq
    C^1(t) := C^1({\tau_{k}}) , \qquad     C^2(t) := C^1({\tau_{k-1}}).
\eeq
    We denote by $C(t)$ the two-dimensional credit rating process defined by
\[
    C(t) = (C^1(t),C^2(t)).
\]
   Therefore the pre-default term structure
depending on $C^1(t)$ is given by the formula
\[
    g(t,u ) = g_{C^1(t)} (t,u ) =
    \I_{\{C^1(t) = 1\}} g_{1} (t,u )
    +
    \ldots
    +
    \I_{\{C^1(t) = K-1\}} g_{K-1} (t,u ) .
\]
We sum up to $K-1$, since the last $K$-th rating corresponds to the
default event
\[
    \tau = \inf \big\{ t > 0 \ : \  C^1(t) = K \big\} .
\]
 It is obvious that each recovery payment depends on the credit rating before
default, i.e.
$$
    \d(t) = \d_{C^2(t)}(t) =
    \I_{\{ C^2(t)  = 1\}}  \d_{1} (t)
    +
    \I_{\{C^2(t) = 2\}} \d_{2} (t)
    +
    \ldots
    +
    \I_{\{C^2(t) = K-1\}} \d_{K-1} (t) ,
$$
where $\d_i$ is a recovery payment connected with the $i$-th rating.
\\
We make a standard assumption on the relationship
between short term spread, recovery and the intensity of migration
into the default state for  defaultable bonds (see e.g.  Jarrow et
al. \cite{JarrLandTurn1997},  Duffie and Singleton
\cite{Duffie1997}).

 \n \textbf{Hypothesis (H1):} \beqa \label{premium}
g_i(t,t) - f(t,t) = \lambda_{i,K}(t)(1-\delta_i(t)) ,\; \quad i
=1, \ldots, K-1, \quad t < T^* \eeqa
 so the
intensity of migration from rating $i$ into the default state $K$ is
equal to the short term credit spread for rating $i$ divided by
one minus recovery from rating $i$. Of course, this does not mean
that the forward rates $f, g$ are strongly linked. It only means
that we cannot specify arbitrarily the intensities of the migration
into the default state $K$ if we have specified $f$, $g$ and the
recovery $\delta$.
 Of course,
\eqref{premium} implies
 \beqa
        \label{eq:HJM1credit_mig}
        g_{C^1(t)}(t,t) = f(t,t) +(1 -  \d_{C^1(t)}(t))
        \la_{C^1(t),K}(t), \quad  t < T^* ,
       \eeqa
 Hypothesis (H1) is natural, which  can be seen from the following facts.
\begin{rem} \label{uw1}
If the price of a defaultable bond  with fractional recovery of
market value is given in a traditional way (see Duffie and
Singleton \cite{Duffie1999}), then it is given by
the intensity proces $\lambda$ and the risk-free short term rate
$r$ in the following way:
\[
    \I_\set{ \tau > t } \h{D}(t,\th) = \I_\set{ \tau > t }
    \EE ( e^{ - \int_t^\th [ r(u) + ( 1 - \d(u) ) \la(u) ] du } | \F_t
    ).
\]
Then, for bounded $\lambda$ and $r$, we have
  \begin{align*} \nonumber
   g_1&(t,t)
\defi  - \lim_{\th \downarrow t } \f{\partial}{\partial\th} \ln \EE ( e^{ -
\int_t^\th [ r(u) + ( 1 - \d(u) ) \la(u) ] du } | \F_t ) = -
\lim_{\th \downarrow t } \f{ \f{\partial}{\partial\th} \EE ( e^{ -
\int_t^\th [ r(u) + ( 1 - \d(u) ) \la(u) ] du } | \F_t  ) }{ \EE (
e^{ - \int_t^\th [ r(u) + ( 1 - \d(u) ) \la(u) ] du } | \F_t  ) }
\\
& \nonumber= - \lim_{\th \downarrow t } \f{ \EE (
\f{\partial}{\partial\th} e^{ - \int_t^\th [ r(u) + ( 1 - \d(u) )
\la(u) ] du } | \F_t  ) }{ \EE ( e^{ - \int_t^\th [ r(u) + ( 1 -
\d(u) ) \la(u) ] du } | \F_t  ) } = \lim_{\th \downarrow t } \f{ \EE
(  [ r_\th + ( 1 - \d_\th ) \la_\th  ]e^{ - \int_t^\th [ r(u) + ( 1
- \d(u) ) \la(u) ] du } | \F_t  ) }{ \EE ( e^{ - \int_t^\th [ r(u) +
( 1 - \d(u) ) \la(u) ] du } | \F_t  )
 }
\\
& \nonumber = r(t) + ( 1 - \d(t) ) \la(t),
\end{align*}
 so \eqref{premium} holds.
\end{rem}
For models with ratings we can make a similar observation.
 We illustrate this in the next propositions where
we assume that the  ex-dividend price of a bond has a natural  form
\eqref{eq2/8} (see e.g. Jakubowski and Niew\k{e}g\l owski
\cite{JakubNiewe2008}) and we demonstrate that this form of prices
implies \eqref{premium}.
\begin{prop} \label{pr1/7}
Let a market of defaultable bonds with fractional recovery of par
value be such that the ex-dividend price $D$ of a bond maturing at
$\theta > 0$ is, on the set $\set{C_t =i }$, $i  \neq K $, equal
to
\begin{eqnarray} \label{eq2/8}
D(t,\theta) {\mathbf{1}}_\set{C_t =i } = {\mathbf{1}}_\set{C_t = i }
\sum_{j=1}^{K-1} \mathbf{E}  \left( e^{ - \int_t^\theta r(v) dv }
p_{i,j}(t,\theta) + \delta_{j} \int_t^\theta e^{ - \int_t^u r(v) dv}
p_{i,j}(t,u) \lambda_{j,K}(u) du  | \mathcal{F}_t \right)
\end{eqnarray}
for  $t < \theta $,  where  $\delta_j$ is the recovery payment for
rating $j$ and $p(t,u)$ is the solution to the (random)
conditional Kolmogorov forward equation
\[
 d p(t,u) = p(t,u ) \Lambda(u) du ; \ \ \ \ \ \ \ p(t,t) = \mathbb{I},
\]
with  the  intensity matrix process $\Lambda$.
Assume that $r$ and
$\Lambda$ are bounded processes. Then
\[
g_i(t,t) =  r(t) + (1 - \delta_i ) \lambda_{i,K}(t) ,
\]
for $i < K $ and $t < \theta $.
\end{prop}
 As we announced in the Introduction, the proof of
this proposition, as well as other proofs, are given in the last
section of the paper.

 It is worth  noticing that the same conclusions
can be drawn for other kinds of recovery. Next, we assume that the
credit migration process and bond price processes have no common
jumps.

 \n \textbf{Hypothesis (H2):}
For the consecutive jump times  $(\tau_k)_{k \geq 0}$ of the credit
migration process
 and for all $\th \in [0,T^*]$ we have
\[
\P( \De B (\tau_k,\th) \neq 0 ) = 0,  \qquad \P( \De D_i
(\tau_k,\th) \neq 0 ) = 0,  \; \forall i =1, \ldots,  K-1 .
\]
\begin{rem} \label{uw2}
For the credit migration process $(C^1(t))_{t
 \in [0,T^{*}]}$ constructed in Bielecki and Rutkowski \cite{Bielecki2002}, \cite{Bielecki2004}
 hypothesis (H2) is fulfilled. This fact follows from Proposition \ref{prop1} below.
\end{rem}
 We also impose the following natural assumption (see
\cite{Bielecki2002}, Blanchet-Scalliet and Jeanblanc
\cite{Blanchet}):

\n \textbf{Hypothesis (H3):} For given filtrations $\mathbb{F}$ and
$\mathbb{G}$, with $\mathbb{F} \subseteq \mathbb{G}$, every
$\mathbb{F}$-local martingale is  a $\mathbb{G}$-local martingale.

In the rest of the paper we assume (H1), (H2) and (H3) for all
semimartingales under consideration.

\subsection{Models without rating migration}

We recall the classical description of such models. The default time
$\tau$ is a $\mathbb{G}$- stopping time, and $ \mathbb{G} =
\mathbb{F} \vee \mathbb{H}$, where $\mathbb{F} = (\F_t)_{t \geq 0}$
and $\mathbb{H} = (\H_t)_{t \geq 0}$ are filtrations generated by
observing the market and observing the default time, i.e.  $\H_t =
\s ( \{\tau \leq u \}: u \leq t) $, respectively. Let $
(H(t))_{t\geq 0}$
  be {\it the default indicator process}, i.e.
  \begin{equation}     \label{eq:def-H}   H(t) =
\I_{\{ \tau \leq t\}} .
   \end{equation}
 We assume that $\tau$ admits an
$\mathbb{F}${\it-martingale intensity} $(\la_t)_{t \geq 0}$ which is
an $\mathbb{F}$-adapted process such that $M_t$ given by the formula
\be\label{eq:intensity-mart-def}
    M_t = H(t) - \int_0^{t \wedge \tau} \la_u du = H(t) - \int_0^{t }(1- H(u)) \la_u du
\ee follows a $\mathbb{G}$-martingale (see  Bielecki and
Rutkowski  \cite{Bielecki2000}).

Since we allow for enlarging the filtration, we need some additional
assumptions under which an $\mathbb{F}$-L\'evy process is a
$\mathbb{G}$-L\'evy process. So we assume hypothesis (H3) holds for
filtrations $\mathbb{F}$ and $\mathbb{G}$.
    In
Bielecki and Rutkowski \cite{Bielecki2000} hypothesis (H3) is also
called condition (M.1) or \emph{Martingale Invariance Property of
 $\mathbb{F}$ with respect to $\mathbb{G}$}. In our case,
if $\tau$ is an $\mathbb{F}$-stopping time, then $\mathbb{G} =
\mathbb{F}$ by definition, so hypothesis (H3) holds. If $\tau$ is
not an $\mathbb{F}$-stopping time, then hypothesis (H3) is
equivalent to conditional independence of the $\s$-fields
$\F_\infty$ and $\G_t$ given $\F_t$ for any $t \in \R_+$ (see Lemma
6.1.1 in Bielecki and Rutkowski
\cite{Bielecki2000}). \\
Moreover, if  $\tau$ is not an $\mathbb{F}$-stopping time, the
assumption that $\tau$ has intensity can be given in an alternative
form, through the assumption that the process $F_t := \P(\tau \leq t
| \F_t)$ is increasing and absolutely continuous w.r.t. Lebesgue
measure. This means that there exists a nonnegative
$\mathbb{F}$-adapted process $f_t$ such that
\[
    F_t := \P(\tau \leq t \mid \F_t) = \int_0^t f_u du .
\]
If we assume that $F_t < 1$, $t \geq 0$,  then we can find an
$\mathbb{F}$-adapted process $(\la_t)_{t \geq 0}$ such that
  \be\label{eq:intensity-interpretation}
   1- F_t = \P( \tau > t \mid \F_t) = e^{- \int_0^t \la_u du} .
\ee
  This process $(\la_t)_{t \geq 0}$ is given by the formula
\be\label{eq:intensity-def}
    \la_t := \f{f_{t}}{1 - F_{t}},
\ee and  one can easily check that $(\la_t)_{t \geq 0}$ is the
$\mathbb{F}$-martingale intensity of $\tau$. Moreover,
\[
    \P( \tau > T \mid \G_t) = \I_\set{ \tau > t } \EE( e^{- \int_t^T \la_u du} | \F_t).
\]
If
  \be\label{eq:cond-independent} \P(\tau \leq t \mid \F_t) = \P(\tau
\leq t \mid \F_\infty) \wsp \forall t \in \R_+  , \ee
  then $F_t$ is
increasing. Bielecki and Rutkowski  \cite{Bielecki2000} show that
this condition (called Condition (F.1a)) is equivalent to hypothesis
(H3) (see Lemma 6.1.2 in \cite{Bielecki2000}).
\begin{example}
Assume that $\tau$  is a random time with density $f > 0$ and
probability distribution $F$  independent of the $\sigma$-field
$\mathbb{F}$. Then the $\mathbb{F}$-intensity of $\tau$ is a
deterministic function given by
    \[
        \la_t = \f{f_t}{1 - F_t} .
    \]
Indeed, independence implies $\P(\tau \leq t | \F_t) = \P(\tau
\leq t) = F_t$. Moreover $M$ given by
\eqref{eq:intensity-mart-def} is a $\mathbb{G}$-martingale, so
$\la_t$ is an $\mathbb{F}$-martingale intensity.
 \end{example}
\begin{example}[Canonical construction of default time,
see section 6.5 in  \cite{Bielecki2000} ] \label{ex:2}

\n If the probability space is sufficiently rich to support a
random variable $U$ uniformly distributed on $[0,1]$ and
independent of $\mathbb{F}$, then for a given
$\mathbb{F}$ adapted nonnegative process $(\la_t)_{t \geq 0}$
satisfying $\int_0^{+ \infty} \la_u du = + \infty $ we can
construct a default time $\tau$ with intensity $(\la_t)_{t \geq
0}$ by the formula
\[
    \tau := \inf \big\{ t \geq 0 : e^{ - \int_0^t \la_u du} \leq U
    \big\}.
\]
One can easily show that $(\la_t)_{t \geq 0}$ is the
$\mathbb{F}$-intensity of $\tau$ (formula
\eqref{eq:intensity-interpretation} holds), and hence also the
$\mathbb{F}$-martingale intensity. Under this construction
\eqref{eq:cond-independent} holds, which implies hypothesis (H3).
\end{example}

We also have
\begin{prop} \label{prop1}
    Let $(X_t)$ be an $\mathbb{F}$-semimartingale, and
$\tau$ a random time given by the canonical construction  with
$(\la_u)_u$ a  strictly positive $\mathbb{F}$-intensity of $\tau$.
Then
    \[
        \P \big(  \De X_\tau \neq 0\big) = 0.
    \]
\end{prop}

 We do not want to assume that $\tau$ is given through the
canonical construction, so we assume (H3) throughout the rest of the
paper. But we emphasize  that if $\tau$ is given through the
canonical construction, then Hypothesis (H3) is redundant.

\begin{rem} \label{uw3}
This model is a special case of the model with rating migration.
Indeed,  taking $K=2$, $C(t)=1+ H(t)$ and the intensity $\lambda$
given by \eqref{eq:intensity-def}, we obtain the previous model (for
details see \cite{Bielecki2002}, p. 396). The conditional generator
of $C$ is of the form
\[
\Lambda(t) =
    \begin{pmatrix}
    - \lambda_t  & \lambda_t \\
    0   & 0
    \end{pmatrix}.
\]
\end{rem}

\section{The HJM conditions for credit risk} \label{sec-HJM}

We consider three types of  recovery payment described in the
previous section and fractional recovery with multiple defaults.
Since we investigate them separately, we use the same notation $D$
for price processes with different recovery payments (so $D$ has
different meanings in different subsections). Let us recall that
all results are obtained under assumptions (A1), (A2)
 for all L\'evy processes considered and (H1),
(H2) and (H3).

\subsection{Models with rating migration}
\subsubsection{Fractional recovery of market value with rating
migration}

    Let us focus on defaultable bonds with fractional recovery of market value
$D(t,\mat)$. This kind of bond pays $1$ unit of cash if default
has not occurred before maturity $\mat$,  i.e., if the default
time satisfies $ \tau
> \mat$, and if the bond defaults before $\mat$ we have recovery payment
at the default time which is a fraction $\d(t)$ of its market value
just before the default time, so the  recovery payment is equal to $
\d (\tau) D(\tau -,\mat)$. Therefore, in the case of rating
migration, the price process of the defaultable bond with credit
migration and fractional recovery of market value should satisfy
\[
    D(\mat,\mat) = \I_{\{ \tau > \mat \}} + \I_{\{ \tau \leq \mat \}} \d_{C^2(\tau)}
    (\tau)
    D_{C^2(\tau)}(\tau-,\mat) \f{B_\mat}{B_\tau} ,
\]
where  $\tau = \inf \{ t > 0 : C^1(t) = K \} $. Hence we postulate
that for $t \leq \mat $
 \beq
    D(t,\mat) &=& \I_{\{ C^1(t) \neq K \}} D_{C^1(t)}(t,\mat) + \I_{\{ C^1(t) = K \}}
    \d_{C^2(\tau)}(\tau)
    D_{C^2(\tau)}(\tau-,\mat)
    \f{B_t}{B_\tau}
    \\
    &=&
    \sum_{i=1}^{K-1}
    \I_{\{ C^1(t) \neq K \}}
    \I_{\{ C^1(t) = i \}}
    D_{i}(t,\mat)
 +
    \sum_{i= 1} ^{K-1}
    \I_{\{ C^1(t) = K \}}
    \I_{\{ C^2(t) = i \}}
    \d_{i}(\tau)
    D_{i}(\tau-,\mat)
    \f{B_t}{B_\tau} .
\eeq
    For $i \neq j $ we define an auxiliary process $H_{i,j} $ by the  formula
    \[
        H_{i,j} (t) = \sum_{0 < u \leq t} H^i (u-) H^j (u), \wsp
        \forall t \in \R_+.
    \]
 This process $H_{i,j}$ counts the number of jumps of  the migration
process $C^1(t)$ from state $i$ to state $j$ up to time t.
 Using the processes $H_i$ and $H_{i,K}$ we can write
$D$ in the form
\begin{equation}     \label{eq:JA1}
    D(t,\mat) = \sum_{i=1}^{K-1} \ \Big( H_i(t) D_i(t,\mat) + H_{i,K}(t) \d_i(\tau)
    D_i(\tau-,\mat)     \f{B_t}{B_\tau} \Big).
 \end{equation}

\begin{thm} \label{thm:frac-market-mig}
 The
processes of discounted  prices of a defaultable bond with  credit
migration and fractional recovery of market value
     are local martingales if and only if the following
    condition holds:

\n     for all $\mat \in [0, T^*]$ and for almost all $\ t \leq
\mat$  on the set $\{ \tau > t \}$
 \beqa
     \label{eq:HJM2credit_mig}
        \int_{t}^\mat \a_{C^1(t)}(t,v) dv =   J_{C^1(t)}
        \bigg( \int_t^\mat \s_{C^1(t)}(t,v) d v
        \bigg)                     +
       \sum_{i=1, i \neq C^1(t)}^{K-1}
       \bigg[
       \f{D_i(t-,\mat)}{D_{C^1(t)}(t-,\mat)}
       -1
       \bigg]
       \la_{C^1(t),i} (t) .
    \eeqa
\end{thm}
 It is worth pointing out that from the proof of
Theorem \ref{thm:frac-market-mig} we obtain immediately
\begin{thm} \label{prop2}
If the processes $D_i, \bar{a}_i, \Lambda$ have continuous
trajectories then the HJM postulate  is equivalent to the
following two conditions:  \eqref{eq:HJM1credit_mig} and the
HJM-type condition \eqref{eq:HJM2credit_mig}.
\end{thm}
 So in this case the HJM postulate implies the HJM-type condition
\eqref{eq:HJM2credit_mig} without assuming hypothesis (H1).
Theorem \ref{prop2} appears for the first time in
\cite{Schmidt05}, but in terms of the derivative of the Laplace
exponent (see points i) and ii) of Theorem
\ref{thm:D-HJM-type-cond}). The same conclusion is true for other
types of recovery with analogous proofs, but we do not
formulate these facts as  separate statements.
\begin{rem} \label{uw3j}
 Theorem \ref{prop2} is not true in the case of
L\'evy noise: see an example  in the last section. Therefore,
Theorem 4.2 in \cite{Ozkan04}, which was proved
 under the stronger assumption than ours (since in the proof the It\^o
formula for processes with values in Hilbert spaces is used) is
not true without some additional assumption.
\end{rem}

\subsubsection{Fractional recovery of Treasury value with rating
migration}

The holder of a defaultable bond with fractional recovery of
Treasury value receives $1$ if there is no default by $\mat$, and
if default has occurred before maturity $\mat$, then  a fixed
amount $\d \in [0,1 ]$ is paid to the bondholder at maturity.
Therefore, since paying $\d$ at maturity $\mat$ is equivalent to
paying $\d B(\tau,\mat)$ at the default time $\tau$, in the case
of fractional recovery of Treasury value with rating migration we
have
\[
    D(\mat,\mat) = \I_{\{ \tau > \mat \}} + \I_{\{ \tau \leq \mat \}}
    \d_{C^2(\tau)},
\]
hence
  \beq
    D(t,\mat) &=& \I_{\{ C^1(t) \neq K \}} D_{C^1(t)}(t,\mat) + \I_{\{ C^1(t) = K \}}
    \d_{C^2(t)} B(t, \mat)
    \\
        &=&
    \sum_{i=1}^{K-1}
    \I_{\{ C^1(t) \neq K \}}
    \I_{\{ C^1(t) = i \}}
    D_{i}(t,\mat) +
    \sum_{i= 1} ^{K-1}
    \I_{\{ C^1(t) = K \}}
    \I_{\{ C^2(t) = i \}}
    \d_{i}
    B(t,\mat)
\eeq
 or, equivalently,
\begin{equation}     \label{eq:JA4}
    D(t,\mat) = \sum_{i=1}^{K-1}\ \Big( H_i(t) D_i(t,\mat) + H_{i,K}(t) \d_i
    B(t,\mat)\ \Big) .
\end{equation}
\begin{thm} \label{thm:frac-treasury-mig}
 The processes of discounted  prices of a defaultable bond with
fractional recovery of Treasury value are local martingales if and
only if the  following  condition holds:

\n     for all $\mat \in [0, T^*]$ and for almost all $\ t \leq
\mat$  on the set $\{ \tau > t \}$
\begin{align}
  \label{eq:HJM2credit_mig_1}
  & \int_{t}^\mat \a_{C^1(t)}(t,u) du =  J_{C^1(t)} \bigg(  \int_t^\mat \s_{C^1(t)}(t,v) d v \bigg)
    + \
    \d_{C^1(t)}
    \bigg[
    \f{B(t-, \mat)}{D_{C^1(t)}(t-, \mat)}
    -1
    \bigg]
    \la_{C^1(t),K}(t)
    \\ \nonumber
  & +
    \sum_{j =1 , j \neq C^1(t)}^{K-1}
    \bigg[
    \f{D_j(t-,\mat)}{D_{C^1(t)}(t-,\mat)} -1
    \bigg]
    \la_{C^1(t),j} (t) .
\end{align}
\end{thm}

\subsubsection{Fractional recovery of par value with rating
migration}

In the case of fractional recovery of par value the holder  of a
defaultable bond receives 1 unit cash if there is no default prior
to maturity and if the bond has defaulted  a fixed fraction $\d$ of
the par value is paid at the default time. Therefore the  payoff  at
maturity has the form
\[
    D(\mat,\mat) = \I_{\{ \tau > \mat \}} + \I_{\{ \tau \leq \mat \}}
    \d_{C^2(t)} \f{B_\mat}{B_\tau},
\]
hence
  \beq
    D(t,\mat) &=& \I_{\{ C^1(t) \neq K \}} D_{C^1(t)}(t,\mat) + \I_{\{ C^1(t) = K \}}
    \d_{C^2(t)}  \f{B_t}{B_\tau}
    \\
        &=&
    \sum_{i=1}^{K-1}
    \I_{\{ C^1(t) \neq K \}}
    \I_{\{ C^1(t) = i \}}
    D_{i}(t,\mat) +
    \sum_{i= 1} ^{K-1}
    \I_{\{ C^1(t) = K \}}
    \I_{\{ C^2(t) = i \}}
    \d_{i} \f{B_t}{B_\tau}
\eeq
 or, equivalently,
\[
    D(t,\mat) = \sum_{i=1}^{K-1}\ \Big( H_i(t) D_i(t,\mat) + H_{i,K}(t) \d_i
    \f{B_t}{B_\tau}\ \Big) .
\]
\begin{thm} \label{thm:par-mig}
 The
processes of discounted  prices of defaultable bond with
fractional recovery of par value are local martingales if and only
if the following condition holds:

\n     for all $\mat \in [0, T^*]$ and for almost all $\ t \leq
\mat$  on the set $\{ \tau > t \}$
 \begin{align}
    \label{eq:HJM2credit_mig_2}
    \int_{t}^\mat \a_{C^1(t)}(t,u) du =& J_{C^1(t)} \bigg(
    \int_t^\mat \s_{C^1(t)}(t,v) d v \bigg)
 + \
    \d_{C^1(t)}
    \bigg[
    \f{1}{D_{C^1(t)}(t-, \mat)}
    -1
    \bigg]
    \la_{C^1(t),K}(t)\\
   &
    \nonumber
     +
    \sum_{j =1 , j \neq C^1(t)}^{K-1}
    \bigg[
    \f{D_j(t-,\mat)}{D_{C^1(t)}(t-,\mat)} -1
    \bigg]
    \la_{C^1(t),j} (t) .
  \end{align}
\end{thm}
\subsubsection{Fractional recovery with multiple defaults and
rating migration }

The HJM models with fractional recovery with multiple defaults
were introduced by Sch{\"o}nbucher \cite{Schonbucher1998}.
 This model describes a situation where a company that
has had to declare default is not liquidated but is restructured.
After  restructuring the firm  may again default in the future.
Sch{\"o}nbucher  investigated defaultable bonds whose face value
is reduced by a fraction $L_{\tau_i}$  at each default time
$\tau_i$, where $L_s$ is an $\mathbb{F}$-predictable process
taking values in $[0,1]$. Therefore, a holder of such a
defaultable bond receives, at maturity $\mat$,
\[
     D^m(\mat,\mat) = \prod_{\tau_i \leq \mat} (1 - L_{\tau_i}) .
\]
If we introduce a process $V_t$ by the formula
\[
     V_t = \prod_{\tau_i \leq t} (1 - L_{\tau_i}) ,
\]
then $ D^m(\mat,\mat)=V_\mat$ and for $ t \leq \mat$,
\be\label{eq:bond-rec-mult}
     D^m(t,\mat) = V_t e^{- \int_t^\mat g_1 (t ,u) du } = V_t D_1(t,
     \mat) .
\ee Moreover, we assume  that $\tau_i$ are jump times of a Cox
process $N_t$ (doubly stochastic Poisson process) with stochastic
intensity process $(\gamma_t)_{t \geq 0}$. It can be shown that
$V_t$ solves the following SDE:
\begin{equation}     \label{eq:schon1}
     d V_t  = - V_{t-} L_t d N_t ,
   \end{equation}
and the process
\begin{equation}     \label{eq:schon2}
     M_t = N_t - \int_0^t \gamma_u du
   \end{equation}
is a $\mathbb{G}$-martingale (Lando \cite{Lando1998}).

In this paper we add a rating migration process to the model. We
assume that the default times are jumps of a Cox process with
intensity $(\gamma_t)_{t \geq 0}$.
Since the company  is restructured after default,  the rating
migration process has no absorbing state and for the rating
migration process $C$ we take  a \cadlag process,  which is an
$\mathbb{F}$-conditional Markov chain with values in the set $\{
1, \ldots, K-1 \}$ without absorbing state. Moreover, we assume
that the process describing fractional losses does not depend on
the credit migration process.
\begin{rem}
     Note that $1- L_t$ can be interpreted as a recovery process and
     therefore we will denote it by $\d(t)$. Thus $\d(t) = 1- L_t$.
\end{rem}
Thus the bond price process should satisfy the following terminal
condition:
\[
     D(\mat,\mat) = V_\mat = \prod_{\tau_i \leq \mat} (1 -
     L_{\tau_i}) = \prod_{\tau_i \leq \mat} \delta_{\tau_i},
\]
and before maturity it should be given by the formula
\[
     D(t,\mat) = V_t D_{C^1(t)} (t,\mat) = V_t \sum_{i=1}^{K-1} H_i(t)D_{i}
     (t,\mat).
\]
\begin{rem}
In this case the filtration $\mathbb{G}$ is specified as
$\mathbb{G} = \mathbb{F} \vee \mathbb{F}^N \vee \mathbb{F}^C $,
i.e. $\G_t = \F_t \vee \F^N_t \vee \F^C_t $,  and hypothesis (H1),
i.e. formula \eqref{premium}, takes the form
$$ \label{eq:HJM1-rec-mult-rating}
         g_{C^1(t)}(t,t) = f(t,t) + (1- \d(t)) \gamma_t  \  . $$
         \end{rem}
\begin{thm}  \label{th:multiple}
The discounted prices of a bond with fractional recovery with
multiple defaults and rating migration are local martingales if and
only if the
     following condition holds: \\
     for all $\mat \in [0, T^*]$ and for almost all $\ t \leq
\mat$  on the set  $\{ V_{t-} > 0\}$
     \begin{align}
         \label{eq:HJM2-rec-mult-rating}
        & \int_{t}^\mat \a_{C^1(t)}(t,v) dv = J_{C^1(t)} \bigg(
         \int_t^\mat \s_{C^1(t)}(t,v) d v \bigg) +
         \sum_{j=1, j \neq {C^1(t)}}^{K-1}
         \bigg[
         \f{D_j(t-,\mat)}{D_{C^1(t)}(t-,\mat)}
         -1
         \bigg]
         \la_{{C^1(t)},j} (t) \ .
      \end{align}
\end{thm}

\subsection{Models without rating
migration}

As we know, taking $K=2$ in the model  with rating migration we
obtain results for models of  defaultable bonds with one credit
rate, so for models without rating migration. To give a clear
picture of markets with defaultable bonds, and for the sake of
completeness, we formulate the HJM drift conditions for these
models:

\begin{thm}
The discounted prices of a defaultable bond are local martingales
if and only if the following condition
    holds
 for all $\mat \in [0, T^*]$ and for almost all $\ t \leq
\mat$  on the set   $\{ \tau > t \}$: \\
1) for
 fractional recovery of market value
     \beqa
       \label{eq:HJM2-rec-market-value}
       \int_{t}^\mat \a_1(t,v) dv &=& J_1 \bigg(  \int_t^\mat \s_1(t,v) d v
       \bigg).
    \eeqa
2) for fractional recovery of Treasury value
 \beqa
    \label{eq:HJM2-rec-treasury}
    \int_{t}^\mat \a_1(t,v) dv &=& J_1 \bigg(     \int_t^\mat \s_1(t,v)
    d v \bigg)
    + \d \bigg( \f{B(t-,\mat)}{D_1(t-,\mat)}   -1  \bigg)
    \la_t.
\eeqa
 3) for fractional recovery of par value
    \beqa
        \label{eq:HJM2-rec-par}
        \int_{t}^\mat \a_1(t,v) dv &=& J_1 \bigg(
        \int_t^\mat \s_1(t,v) d v \bigg)
         + \d \bigg(  \f{1}{D_1(t- , \mat)} -1 \bigg)
        \la_t .
    \eeqa
\end{thm}

\begin{thm}
   The discounted prices of a defaultable bond with multiple
defaults and fractional recovery are local martingales if and only
if the following condition holds: \\
  for all $\mat \in [0, T^*]$ and for almost all $\ t \leq
\mat$  on the set  $\{ V_{t-} > 0\}$,
    \beqa
         \label{eq:HJM2-rec-mult}
        \int_{t}^\mat \a_1(t,v) dv &=& J_1 \bigg(  \int_t^\mat
\s_1(t,v)
     d v \bigg) .
     \eeqa
\end{thm}

\section{The HJM condition in terms of the derivative of the Laplace
exponent}
 If the derivative of the Laplace exponent exists,
then the HJM conditions have simpler forms. To obtain these forms
we use some facts on such derivatives, including
\begin{lem} \label{lem1}
    Let $G$ be a functional defined on an open subset $B_1$ of $U$, of the form
    \[
        G(x) = \int_U \big( e^{-\inh{x,y}} -1 + \I_{\normh{y} \leq 1}(y) \inh{x,y} \big) \nu (dy),
    \]
where $\nu$ is a L\'evy measure which has exponential moments
\begin{equation}  \label{mom-dod}
    \int_\set{ \normh{y} >1 }
    e^\inh{ c,y}
    \nu(dy)< +
    \infty
\end{equation}
for all $c \in B_1$.
 Then
 $G$ is differentiable at each $x \in B_1$ and
    \[
        D G(x) = - \int_U \big( e^{- \inh{x,y}}  - \I_{\normh{y} \leq 1}(y) \big) y \ \nu (d
        y).
    \]
\end{lem}
 The proof  is straightforward. We use the existence of exponential moments
 of the form \eqref{mom-dod} for all $c \in B_1$.

 Hence, for models with ratings, after
straightforward calculations we obtain the HJM  conditions  in
terms of the derivatives of the Laplace exponents $J_i$,
$i=1,2,\ldots,k-1$.
\begin{thm} \label{thm:D-HJM-type-cond}
Assume that for $i=1,2,\ldots,K-1$
 \begin{equation}
    \label{eq:n}
  \int_{\{|y| \geq 1\}} e^{- \inh{u, y }}\, \nu_i(dy) <\infty
\end{equation}
 for all $u$ from some neighborhood of
the set in  which $\int_t^{\theta} \sigma_i(t,v) dv$ takes values.
 Then \\
  i)  Condition
    (\ref{eq:HJM2credit_mig}) for fractional recovery of market
    value and condition (\ref{eq:HJM2-rec-mult-rating}) for fractional recovery with multiple defaults have the form
    \beq
        && \a_{C^1(t)}(t,\mat )  =  \binh{ D J_{C^1(t)} \bigg( \int_0^\mat \s_{C^1(t)}(t,v) d v \bigg), \s_{C^1(t)}(t,\mat) }
        \\
        &&
        +
        \sum_{i= 1,i \neq C^1(t)}^{K-1}
       \la_{C^1(t),i} (t)
       \Big(
       g_{C^1(t)}(t-,\mat)
       -g_i(t-,\mat)
       \Big)
       e^{ \int_t^\mat (g_{C^1(t)}(t-,u) - g_i(t-,u)) du }        .
    \eeq
   ii)  Condition
    (\ref{eq:HJM2credit_mig_1}) for fractional recovery of Treasury value has the form
    \beq
        &&
        \a_{C^1(t)}(t,\mat )  =  \binh{ D J_{C^1(t)} \bigg( \int_0^\mat \s_{C^1(t)}(t,v) d v \bigg), \s_{C^1(t)}(t,\mat) }
            \\
        &&
        +
        \sum_{i= 1, i \neq C^1(t)}^{K-1}
       \la_{C^1(t),i} (t)
       \Big(
       g_{C^1(t)}(t-,\mat)
       -g_i(t-,\mat)
       \Big)
       e^{ \int_t^\mat (g_{C^1(t)}(t-,u) - g_i(t-,u) )du }
       \\
       &&  + \
        \d_{C^1(t)} \la_{C^1(t),K}
        \Big(
        g_{C^1(t)}(t-,\mat)
        -f(t-,\mat)
        \Big)
        e^{\int_t^\mat (g_{C^1(t)}(t-,u) - f(t-,u)) du } .
    \eeq
  iii)  Condition
    (\ref{eq:HJM2credit_mig_2}) for   fractional recovery of par value  has the form
    \beq
        && \a_{C^1(t)}(t,\mat )  = \binh{ D J_{C^1(t)} \bigg( \int_0^\mat \s_{C^1(t)}(t,v) d v \bigg), \s_{C^1(t)}(t,\mat) }
        \\
        &&
        +
        \sum_{i =1, i \neq C^1(t)}^{K-1}
       \la_{C^1(t),i} (t)
       \Big(
       g_{C^1(t)}(t-,\mat)
       -g_i(t-,\mat)
       \Big)
       e^{ \int_t^\mat (g_{C^1(t)}(t-,u) - g_i(t-,u)) du } \\
       && + \
        \d_{C^1(t)} \la_{C^1(t),K} g_{C^1(t)}(t-,\mat)
        e^{\int_t^\mat g_{C^1(t)}(t-,u) du } .
    \eeq
\end{thm}
 For  infinite dimensional  Brownian motion points i) and ii) of
Theorem \ref{thm:D-HJM-type-cond} were proved in \cite{Schmidt05}.
As a simple consequence of Theorem \ref{thm:D-HJM-type-cond} we
obtain
 \begin{cor} \label{uw7}
 Under the assumption of Theorem \ref{thm:D-HJM-type-cond} on $J_1$, the conditions
for models without ratings have a simpler form, namely: \\
  i)  Condition
    (\ref{eq:HJM2-rec-market-value}) for  fractional recovery of market value has the form
    \beq
        \a_1(t,\mat)  &=& \binh{ D J_1 \bigg(     \int_0^\mat \s_1(t,v) d v \bigg)  , \s_1(t,\mat)
        } .
    \eeq
  ii) Condition (\ref{eq:HJM2-rec-treasury}) for   fractional recovery of Treasury value has the form
    \beq
    \a_1(t,\mat ) &=& \binh{D J_1 \bigg(     \int_0^\mat \s_1(t,v) d v \bigg), \s_1(t,\mat)   }
    \\
    &&+
        \d \la_t
        \Big(
        g_1(t-,\mat)
        -f(t-,\mat)
        \Big)
        e^{ \int_t^\mat (g_1(t-,u) - f(t-,u)) du } .
    \eeq
 iii)   Condition (\ref{eq:HJM2-rec-par}) for fractional recovery of par value has the form
    \beq
       \a_1(t,\mat ) &=& \binh{ D J_1 \bigg(     \int_0^\mat \s_1(t,v) d v \bigg), \s_1(t,\mat)
       }
       + \d \la_t g_1(t-,\mat)e^{ \int_t^\mat g_1(t-,u) du }.
    \eeq
    \end{cor}

\section{Comparison of consistency conditions  and HJM conditions} \label{sect5}

In this section we investigate the relationships between
consistency conditions formulated by Bielecki and Rutkowski and
the HJM conditions introduced in the previous section.  The papers
\cite{Bielecki2000} and \cite{Ozkan2003} provide an
exogenously specified term structure of defaultable forward rates
corresponding to a given finite collection of credit ratings and
then the authors look for an arbitrage free model that supports these
objects. They are interested in  whether there exists a rating
migration process such that defaultable bond price processes have
prespecified defaultable forward rates. They require this
system of prices to be consistent in the sense that the discounted
defaultable price processes are local martingales under an
appropriately chosen equivalent probability measure. They provide
conditions for the intensity matrix processes which guarantee this
kind of {"consistency"}, which means that the HJM postulate is
satisfied. Hence the {"consistency condition"} should be related
in some way to the HJM drift-condition derived in the previous
section. Now we investigate this relation.
  First, note
that the consistency conditions in Bielecki and Rutkowski
\cite{Bielecki2000} and Eberlein and \"{O}zkan \cite{Ozkan2003} are
given under a real-world probability measure, and our HJM
conditions are related to a risk-neutral world.
 So we formulate  consistency conditions assuming that we are in a
 risk-neutral world.
We start with the  case of fractional recovery of market value
with rating migration. We say that the  consistency condition
(cf. \cite{Bielecki2002}, \cite{Ozkan2003}) holds if the
equalities
  \beqa \label{eq:cons-market-value}
    && \sum_{i=1, i \neq C^1(t)}^{K-1}
    \bigg[
    \big(
    D_i(t-,\mat)
    -D_{C^1(t)}(t-,\mat)
    \big)
    \la_{C^1(t),i} (t)   \bigg] + \\
    && \nonumber +
    ( \d_{C^1(t)}(t)D_{C^1(t)}(t-,\mat) -D_{C^1(t)}(t-,\mat)
    )\la_{C^1(t),K}(t) \\
    && \nonumber
        +
    \left(g_{C^1(t)}(t,t) - f(t,t) + \bar{a}_{C^1(t) }(t,\mat) \right)D_{C^1(t)}(t-, \mat)    =0
\eeqa
 are satisfied on the set $\{ C^1(t) \neq K \}$  for all $ \mat \in [0,T^*]$  and all
$t \leq \mat $.     We recall that  $\bar{a}_i(t,\mat)$ is defined by \eqref{eq:drift-ai}.

The following theorem states that the consistency condition and HJM
type condition are equivalent under Hypothesis  (H1).
\begin{thm}\label{thm:cons-HJM-type-cond}
Assume that Hypothesis (H1) holds. For defaultable bonds with
credit migration and  fractional recovery of market value the
consistency condition \eqref{eq:cons-market-value} holds if and
only if the HJM type condition \eqref{eq:HJM2credit_mig} holds.
\end{thm}
This theorem allows us to generalize, under (H1), the results of
\cite{Bielecki2000} and \cite{Ozkan2003} to the case of infinite
dimensional L\'evy processes.
\begin{cor}\label{wn:cons-HJM-type-cond-a}
Assume that Hypothesis (H1) holds. If the consistency condition
\eqref{eq:cons-market-value} holds, then the market is arbitrage
free.
\end{cor}
Moreover, we also have an inverse implication:
\begin{cor}\label{wn:cons-HJM-type-cond-b}
Assume that Hypothesis (H1) holds. If the HJM postulate is satisfied,
then the consistency condition \eqref{eq:cons-market-value} holds.
\end{cor}

 In the case of other kinds of recoveries we have a similar
situation although consistency conditions have a slightly different
form. For fractional recovery of  treasury  value with rating
migration the consistency condition is of the form
  \beqa  \label{eq:cons-treasury}
     && \sum_{j =1,j \neq C^1(t)}^{K-1}
    \bigg[
             (
     D_j(t-,\mat) -D_{C^1(t)}(t-,\mat)
     )
     \la_{C^1(t),j} (t) \bigg] \\ \nonumber
     && +
     (
     \d_{C^1(t)}
     B(t-, \mat)
     -
     D_{C^1(t)}(t-, \mat)
     )
     \la_{C^1(t),K}(t)
     \\ \nonumber
     && +
         \left(g_{C^1(t)}(t,t) - f(t,t) + \bar{a}_{C^1(t) }(t,\mat) \right) D_{C^1(t)}(t-, \mat)
             =0
\eeqa on the set $\{ C^1(t) \neq K \}$  for all $ \mat \in
[0,T^*]$ and all $t \leq \mat $. \\
  In the
   case of fractional recovery of par value with rating
migration the consistency condition has the form
  \beqa \label{eq:cons-par-value}
     && \sum_{i=1, i \neq C^1(t)}^{K-1}
    \bigg[
     \big(
     D_i(t-,\mat)
     -D_{C^1(t)}(t-,\mat)
     \big)
     \la_{C^1(t),i} (t) \bigg] + \\ \nonumber
     &&
     ( \d_{C^1(t)}(t) -D_{C^1(t)}(t-,\mat) )\la_{C^1(t),K}(t)
        +
         \left(g_{C^1(t)}(t,t) - f(t,t) + \bar{a}_{C^1(t) }(t,\mat) \right)D_{C^1(t)}(t-, \mat)
          =0 .
\eeqa In the case of fractional recovery with multiple defaults
with rating migration the consistency condition has the form
\beqa \label{eq:cons-mult-defaults}
     && \sum_{i=1, i \neq C^1(t)}^{K-1}
    \bigg[
     \big(
     D_i(t-,\mat)
     -D_{C^1(t)}(t-,\mat)
     \big)
     \la_{C^1(t),i} (t)  \bigg] + \\ \nonumber
     && +
     ( \d_{t}D_{C^1(t)}(t-,\mat)  - D_{C^1(t)}(t-,\mat)  ) \la_{t}
         \left(g_{C^1(t)}(t,t) - f(t,t) + \bar{a}_{C^1(t) }(t,\mat) \right)D_{C^1(t)}(t-, \mat)
          =0 .
\eeqa
  Arguing as  in Theorem \ref{thm:cons-HJM-type-cond} we obtain
\begin{thm}\label{eq:cons-HJM-type-cond2}
    Assume that Hypothesis (H1) holds. Then\\
i)
  The HJM-type condition \eqref{eq:HJM2credit_mig_1} for defaultable bonds with  credit migration and fractional
recovery of  treasury value is equivalent to the consistency
condition
(\ref{eq:cons-treasury}) .\\
ii)
  The HJM-type condition \eqref{eq:HJM2credit_mig_2} for defaultable bonds with  credit migration and fractional
recovery of   par value  is equivalent to the consistency condition
(\ref{eq:cons-par-value}). \\
iii)
  The HJM-type condition \eqref{eq:HJM2-rec-mult-rating} for defaultable bonds with  credit migration and multiple
defaults with fractional recovery   is equivalent to the consistency condition
(\ref{eq:cons-mult-defaults}).
\end{thm}
It is worth  noticing that we can formulate and prove results
analogous to those in Corollaries \ref{wn:cons-HJM-type-cond-a} and
\ref{wn:cons-HJM-type-cond-b} for all kinds of recoveries.
 We have just shown that if Hypothesis (H1) holds then HJM type
conditions are equivalent to equalities known as consistency
conditions. We should stress however that Bielecki and Rutkowski
\cite{Bielecki2000} and also Eberlein and \"{O}zkan \cite{Ozkan2003}
treated these conditions as conditions on the intensity matrix process.
They show that if we specify a real-world
dynamics of defaultable forward rates, and then construct a migration
process with intensity matrix satisfying  the {"consistency condition"},
 then we obtain an arbitrage free model of defaultable bonds.
Note that if we specify the transition intensity matrix then we cannot
specify the  volatilities arbitrarily. More precisely, this means that
if the transition intensity matrix is  specified and we are in
an arbitrage free framework then we calculate prices (conditional
prices, i.e. on the sets $\set{ C_t = i }$) and then extract from
them the defaultable forward rates, to finally get the
volatilities. In our  framework hypothesis (H1) gives intensities
of migration to the default state, and by specifying the volatilities and
then choosing $\lambda_{i,j}$ for $j \neq K$ in such a way that
the HJM condition holds we have specified a risk-neutral dynamics of
defaultable forward rates.

\section{Proofs}
In what follows we use the following facts from Bielecki and
Rutkowski \cite{Bielecki2002} (see also \cite{Rogers},
\cite{Brandt}). \\
 If
 $H_i (t) = \I_{\{ i \} } ( C^1(t)) $, then

\[
    M_i(t) = H_i (t) - \int_{0}^t \la_{C^1(u),i} (u) du
\]
is a $\mathbb{G}$-martingale. Recall that
    \[
        H_{i,j} (t) = \sum_{0 < u \leq t} H^i (u-) H^j (u), \wsp
        \forall t \in \R_+.
    \]
    For arbitrary $i,j \in \mathcal{K}, i \neq j$, the processes
    \[
        M_{i,j}(t)
        =
        H_{i,j}(t) - \int_{0}^t \la_{i,j}(u) H_i (u)  du
        =
        H_{i,j}(t) - \int_{0}^t \la_{C^1(u),j}(u) H_i (u)  du
    \]
    and
    \[
        M_{K}(t)
        = H_K(t) - \int_{0}^t \sum_{i=1}^{K-1} \la_{i,K} H_i(u) du
        = H_K(t) - \int_{0}^t \la_{C^1(u),K} ( 1 - H_K(u) ) du
    \]
    are  $\mathbb{G}$-martingales.

Using these facts and assumption (H2) we obtain very useful
representations of $d(H_i(t) \f{D_i(t,\mat)}{B_t})$:
 \beqa
       \nonumber
   && d \Big( H_i(t) \f{D_i(t,\mat)}{B_t} \Big)
    =
    d ( H_i(t) )
    \f{D_i(t-,\mat)}{B_t}
    +
    H_i(t-)
    d
    \Big(
    \f{D_i(t,\mat)}{B_t}
    \Big)
    +
    d
    \underbrace{
    \Big[
    H_i(\cdot),
    \f{D_i(\cdot,\mat)}{B_{\cdot}}
    \Big]^c_t
    }_{=0}
 \\ &&
     \label{eq:JA2}
      +
    \underbrace{
    \De
    H_i(t)
    \De
    \f{D_i(t,\mat)}{B_{t}}
    }_{=0}  =    d ( H_i(t) )
    \f{D_i(t-,\mat)}{B_t}
    +
    H_i(t-)
    d
    \Big(
    \f{D_i(t,\mat)}{B_t}
    \Big) .
\eeqa
 Since the process
$    M_i(t) = H_i (t) - \int_{0}^t \la_{C^1(u),i} (u) du $ is a
$\mathbb{G}$-martingale, using \eqref{eq:JA2} we obtain
\begin{align} \nonumber
   & d \Big( H_i(t) \f{D_i(t,\mat)}{B_t} \Big)
   =
    \f{D_i(t-,\mat)}{B_t}
    \bigg(
    d M_i(t)
    +
    \la_{C^1(t),i} (t) dt +
    H_i(t-)(g_i(t,t) - f(t,t) + \bar{a}_i (t, \mat)) dt
     \\ \label{eq:JA3}
    & +
    H_i(t-)
       \int_\Hil
       \bigg[
       e^{
      \inh{ \int_s^\mat \s_i(s,v) dv ,
         y
       }} -1
       \bigg] (\mu_i(dt, dy) - dt \nu_i (dy))
     -
    H_i(t-)
       \inh{ \int_s^\mat \s_i(s,v) dv
        ,
        dW_i(t)
        }
    \bigg) .
 \end{align}

{ {\bf Proof of Proposition \ref{pr1/7}.} \n Let
$D_{i} (t , \theta )$ be the price process on the set $\set{C_t =
i }$, i.e.
\[
D_{i} (t , \theta ) := D(t,\theta) {\mathbf{1}}_\set{C_t =i } =
\sum_{j=1}^{K-1} \mathbf{E}  \left( e^{ - \int_t^\theta r_v dv }
p_{i,j}(t,\theta) + \delta_{j} \int_t^\theta e^{ - \int_t^u r_v
dv} p_{i,j}(t,u) \lambda_{j,K}(u) du  | \mathcal{F}_t \right) .
\]
Obviously,
\begin{eqnarray} \label{eq1/8}
g_i(t,t) := - \lim_{ \theta \downarrow t}\frac{\partial}{ \partial
\theta} \ln D_i(t,\theta)  = - \lim_{ \theta \downarrow t}
\dfrac{ \frac{\partial}{
\partial \theta} D_i(t,\theta) }{ D_i(t,\theta) } .
\end{eqnarray}
First  note,  by definition of $D_i$, that
\[
\lim_{ \theta \downarrow t} D_i(t,\theta) = 1
\]
for $\ i \neq K$. Let
\[
A_{i,j} (t,\theta) := e^{ - \int_t^\theta r_v dv }
p_{i,j}(t,\theta), \qquad B_{i,j} (t,\theta) := \delta_{j}
\int_t^\theta e^{ - \int_t^u r_v dv} p_{i,j}(t,u) \lambda_{j,K}(u)
du .
\]
Then
\begin{eqnarray*}
\frac{\partial}{ \partial \theta} D_i(t,\theta) &=&
\sum_{j=1}^{K-1} \mathbf{E} \left( \frac{\partial}{ \partial
\theta}A_{i,j} (t,\theta) + \frac{\partial}{ \partial
\theta}B_{i,j} (t,\theta) | \mathcal{F}_t \right)
\end{eqnarray*}
since $r$ and $\Lambda$ are bounded processes. Next
we calculate the derivatives using the conditional Kolmogorov forward
equation for $P(t,\theta)$, and letting $\theta \downarrow t$
we obtain
\[
\lim_{\theta \downarrow t } \frac{ \partial }{ \partial \theta }
A_{i,j} (t,\theta ) = -r_t  \varrho_{i, j } + \sum_{k=1}^{K}
\varrho_{i, k } \lambda_{k,j} (t) = -r_t  \varrho_{i, j } +
\lambda_{i,j} (t), \quad \lim_{\theta \downarrow t } \frac{
\partial }{
\partial \theta } B_{i,j} (t,\theta ) = \delta_j  \varrho_{i, j }
\lambda_{j,K}(t) ,
\]
where $\varrho_{i,j}$ denotes the Kronecker delta.
Therefore, by passing to the limit inside the conditional
expectation we have
\[
\lim_{ \theta \downarrow t } \frac{\partial }{ \partial \theta }
D_{i}(t,\theta) = -r_t + \sum_{j=1}^{K-1} \lambda_{i,j} (t) +
\delta_i \lambda_{i,K}(t) =  -r_t - (1 -  \delta_i)
\lambda_{i,K}(t),
\]
since $\sum_{j=1}^{K-1} \lambda_{i,j} (t) = - \lambda_{i,K}(t) $.
This and \eqref{eq1/8} complete the proof. \kd }

{\bf Proof of Proposition \ref{prop1}.} Let $\sigma$ be a jump time
of $(X_t)$, and $\tau$ be a random time given by the canonical
construction. Since $e^{-\int_0^\sigma \la_u du} $ is
$\F_\s-$measurable and  $U$ is uniformly distributed on $[0, 1]$
and independent of $\F_\s$ we obtain
 \beq
    \P( \tau  = \s )
    &=&
    \P\Big( e^{-\int_0^\tau \la_u du} = e^{-\int_0^\s \la_u du} \Big)
    =
    \P\Big( U = e^{-\int_0^\s \la_u du} \Big)
    \\
    &=&
    \EE\Big( \EE \Big( \I_\set{ U = e^{-\int_0^\s \la_u du}  }  \Big|  \F_\s \Big)\Big)
    =
    \EE\Big( \EE \Big( \I_\set{ U = x }  \Big) \Big|_{x= e^{-\int_0^\s \la_u du} } \Big)
    =0.
\eeq
 Since a semimartingale is a \cadlag\ process,
the set of jump times of $X$ is countable, so
\[
    \P \big(  \De X_\tau \neq 0\big)
    =
    \P \bigg( \bigcup_{n \geq 1} \{ \tau = \s_n \} \bigg)
    \leq
    \sum_{n=0}^\infty
    \P (  \tau = \s_n )
    =0.
\]
\kd

{ \bf Proof of Theorem \ref{thm:frac-market-mig}. } By
\eqref{eq:JA1}, \beqa
        \label{eq:HJM1credit_mig-n}
    d  \Big( \f{D(t,\mat)}{B_t} \Big)
    =
    \sum_{i=1}^{K-1}\ \Big(  d \Big( H_i(t) \f{D_i(t,\mat)}{B_t} \Big) + d \Big( H_{i,K}(t) \d_i(\tau)
      \f{D_i(\tau-,\mat)}{B_\tau} \Big) \Big) .
 \eeqa
 The first term in this sum is given by \eqref{eq:JA3}, and the second has the
form
\[
d \Big( H_{i,K}(t) \d_i(\tau)
      \f{D_i(\tau - ,\mat)}{B_\tau} \Big)
=
 \d_i(t)
      \f{D_i(t-,\mat)}{B_{t}}
  d \Big( H_{i,K}(t) \Big) .
\]
Since the process $ M_{i,K} (t)= H_{i,K}(t) - \int_{0}^t
\la_{i,K}(u) H_i (u)  du $ is a $\mathbb{G}$-martingale, we have
 \beq
 \d_i(t)
      \f{D_i(t-,\mat)}{B_t}
  d \Big( H_{i,K}(t) \Big)
&=&
  \f{D_i(t-,\mat)}{B_t}
   \d_i(t)
  d M_{i,K} (t)
  +
   \f{D_i(t-,\mat)}{B_t}
   \d_i(t) \la_{i,K}(t) H_i (t)
   dt .
\eeq
 Combining these results we see that the drift term $I$ of
 \eqref{eq:HJM1credit_mig-n} is given by
\begin{align} \nonumber
   I(t,\mat) = & \int_0^t  \sum_{i=1}^{K-1}
    H_i(s) \f{D_i(s-,\mat)}{B_s} (g_i(s,s) - f(s,s) + \bar{a}_i (s, \mat)+ \d_i(s) \la_{i,K}(s)) ds
    \\
    &
     +
     \int_0^t
    \sum_{i=1}^{K-1}
    \f{D_i(s-,\mat)}{B_s}
    \la_{C^1(s),i} (s) ds \\  \nonumber
 &  = \int_0^t  (1- H_K(s))
    \f{D_{C^1(s)}(s-,\mat)}{B_s} (g_{C^1(s)}(s,s) - f(s,s) + \bar{a}_{C^1(s)}(s, \mat)     +
    \d_{C^1(s)}(s) \la_{i,K}(s)) ds
     \\    \label{eq:drift-market-value} &
     + \int_0^t
    \sum_{i=1}^{K-1}
    \f{D_i(s-,\mat)}{B_s}
    \la_{C^1(s),i} (s) ds .
 \end{align}
  Since $D_{C^1(s)} > 0$ and
\begin{align*}
    & \sum_{i=1}^{K-1}
    \f{D_i(s-,\mat)}{D_{C^1(s)}(s-,\mat)}
    \la_{C^1(s),i} (s) =
    \sum_{ i=1, i \neq C^1(s)}^{K-1}
    \f{D_i(s-,\mat)}{D_{C^1(s)}(s-,\mat)}
    \la_{C^1(s),i} (s)
    +
    \la_{C^1(s),C^1(s)} (s)  \\
    & =
    \sum_{i=1, i \neq C^1(s)}^{K-1}
    \bigg[
    \f{D_i(s-,\mat)}{D_{C^1(s)}(s-,\mat)}
    -1
    \bigg]
    \la_{C^1(s),i} (s)
    -
    \la_{C^1(s),K} (s),
 \end{align*}
 we can split $I$ into two parts: $I_1(s)$, independent of $\mat$, and
 $I_2(s,\mat)$,
depending on both $s$ and $\mat$, i.e.
\[
  I(t,\mat) =   \int_0^t (1-H_K(s)) \f{D_{C^1(s)}(s-,\mat)}{B_s} (I_1(s) + I_2(s,\mat)) ds,
\]
where
\[
    I_1(s)
    =
    \bigg(
    g_{C^1(s)}(s,s) - f(s,s) - ( 1 -\d_{C^1(s)}(s) )\la_{C^1(s),K}(s)
    \bigg) ,
\]
and
\[
  I_2(s,\mat) =
       \bigg(  {\bar{a}_{C^1(s)}}(s, \mat) +
       \sum_{i=1, i \neq C^1(s)}^{K-1}
    \bigg[
    \f{D_i(s-,\mat)}{D_{C^1(s)}(s-,\mat)}
    -1
    \bigg]
    \la_{C^1(s),i} (s) \bigg)  .
\]
If (\ref{eq:HJM1credit_mig}) and (\ref{eq:HJM2credit_mig}) hold,
then
 $I_1(t) = 0$ and $I_2(t,\mat)=0$, which implies that the drift term $I(\cdot, \mat)$
 vanishes for any $\mat$, so the HJM postulate is satisfied.

\n Conversely, if the drift term
 vanishes for each  $\mat \in [0,T^*]$, then  on the set $\{ C^1(t) \neq K \} = \{ \tau > t \}$,
\begin{equation}\label{eq:drift-vanish}
 I_1(t) + I_2(t, \mat) = 0 \qquad
\text{ for} \  \text{ almost all } t \in [0, \mat] ,
\end{equation}
since $\frac{D_{C^1 ( s ) }(s-,\mat ) }{  B_s } > 0 $.
 From (\ref{eq:HJM1credit_mig}) we obtain $I_1(t)
= 0$. Therefore $I_2(t,\mat)=0$ for almost all $ t \in [0, \mat]$,
which is equivalent to (\ref{eq:HJM2credit_mig}). \kd

    {\bf Proof of Theorem \ref{prop2} . \ }
 Since $I_1$ and $I_2(\cdot,\mat )$ are right
continuous, vanishing of the drift term $I$ for each $\mat \in
[0,T^*]$ implies that
\[
 I_1(t) + I_2(t, \mat) = 0
 \wsp
 \forall t < \mat.
\]
Since  $I_2(\mat,\mat) = 0$ by definition, we obtain
$I_1(\mat-) = 0$ for $ \theta  < T^* $. Hence we deduce that
$I_1(t) = 0$ for $ t < T^* $, which is equivalent to
(\ref{eq:HJM1credit_mig}).

{\bf Proof of Remark \ref{uw3j} } The reason why we could not
omit the assumption (\ref{eq:HJM1credit_mig}) in  Theorem
\ref{thm:frac-market-mig}, so  Theorem \ref{prop2} is not true
without the continuity assumption, is  that \eqref{eq:drift-vanish}
does not imply $ I_1(\mat) + I_2 (\mat , \mat ) = 0$ for $ \theta
< T^* $ a.s., which gives $I_1(\mat) =
0 $ for $ \theta < T^* $ a.s., i.e. (\ref{eq:HJM1credit_mig}). \\
An example that shows that this implication does not hold is
obtained by taking as $\lambda_{i,K}$, $i=1,2,\ldots,K-1$, some
\cadlag processes such that
\[
\mathbf{P}( \exists \theta  \in [0, T^* ]: \quad
|\lambda_{C^{1}(\theta),K}(\theta) -
\lambda_{C^{1}(\theta),K}(\theta-) |
>  0  ) > 0
\]
and then  defining
\[
g_{C^{1}(t) }(t,t) := f(t,t) + (1 - \delta_{ C^{1}(t) })
\lambda_{C^{1}(t), K }(t-) .
\]
We note  that this choice of $g_i$ gives
\[
I_1(\theta) =  (1 - \delta_{ C^{1}(\theta) })
(\lambda_{C^{1}(\theta), K }(\theta-) -  \lambda_{C^{1}(\theta), K
}(\theta)) ,
\]
so
\[
\mathbf{P} \Big( \exists \theta  \in [0, T^* ]: \quad |
I_1(\theta) |
>  0  \Big) = \mathbf{P}\Big( \exists \theta  \in [0, T^* ]: \quad
|\lambda_{C^{1}(\theta),K}(\theta) -
\lambda_{C^{1}(\theta),K}(\theta-) |
>  0  \Big) > 0
\]
even though we have $I_1(s) = 0$ \ $ds \times d \mathbf{P}$ almost surely.

 {\bf Proof of Theorem \ref{thm:frac-treasury-mig}. \ } By
\eqref{eq:JA4}, the discounted value of a defaultable bond with
fractional recovery of  Treasury value equals
\begin{equation}     \label{eq:JA5}
    d  \Big(  \f{D(t,\mat)}{B_t}\Big)  = \sum_{i=1}^{K-1} \
    \Big(   d  \Big( H_i(t) \f{D_i(t,\mat)}{B_t}\Big) +
      d  \Big( H_{i,K}(t) \d_i
    \f{B(t,\mat)}{B_t}\Big)\ \Big) .
\end{equation}
 By integration by parts we have
 \beq
 &&  d \bigg(
H_{i,K}(t)  \d_i
    \f{B(t,\mat)}{B_t}
    \bigg)
    =
    \d_i
    \f{B(t-,\mat)}{B_t}
    \bigg(
    d M_{i,K}(t)
    +
    \big(
    \la_{i,K}(t) H_i(t) +
    \bar{a}(t,\mat) H_{i,K}(t-) \big) dt   \\
    && +
    H_{i,K}(t-)
    \Big(
       \int_\Hil
       \bigg[
       e^{
       -
       \inh{\int_t^\mat \s(t,v) dv ,
         y
       }} -1
       \bigg] (\mu(dt, dy) - dt \nu (dy))
       -
        \inh{
        \int_t^\mat \s(t,v) dv
        ,
        dW(t)
        }
    \Big)
    \bigg) .
\eeq Together with \eqref{eq:JA3} this implies that the drift term
$I$ in \eqref{eq:JA5} is given by
$$
   I =  I_1 + I_2 + I_3 ,
$$
 where
 \begin{align*}
   I_1 =&  \sum_{i=1}^{K-1}
    H_i(t-)
    \f{D_i(t-,\mat)}{B_t}
    \bigg(
    (g_i(t,t) - f(t,t) + {\bar{a}_i}(t, \mat))
    +
    \d_i
    \f{B(t-, \mat)}{D_i(t-, \mat)}
    \la_{i,K}(t)
    \bigg) dt ,
    \\
   I_2 =&  \sum_{j=1}^{K-1}
    \f{D_j(t-,\mat)}{B_t}
    \la_{C^1(t),j} (t) dt
   , \\
   I_3 =&  \f{B(t-, \mat)}{B_t}
    \bar{a}(t,\mat)
    \sum_{i=1}^{K-1}
    \d_i  H_{i,K}(t-) dt
    =
    H_K(t)
    \bigg(
    \f{B(t-, \mat)}{B_t}
    \bar{a}(t,\mat)
    \sum_{i=1}^{K-1}
    \d_i  \I_\set{  C^2 (t) = i } dt
    \bigg) .
  \end{align*}
Now $I_3=0$, because the HJM-type condition for default-free bonds
holds (condition (\ref{eq:HJM-type-condition-J})). Moreover
 \begin{align} \nonumber
 I_2 =&  \sum_{j=1}^{K-1}
    \f{D_j(t-,\mat)}{B_t}
    \la_{C^1(t),j} (t) dt
    =
    \sum_{j=1}^{K-1}
    \f{D_j(t-,\mat)}{B_t}
    \sum_{i=1}^{K-1}
    H_i(t)
    \la_{i,j} (t) dt
    \\  \nonumber
    & =
    \sum_{i=1}^{K-1}
    H_i(t)
    \f{D_i(t-,\mat)}{B_t}
    \bigg(
    \sum_{j=1,j \neq i}^{K-1}
    \f{D_j(t-,\mat)}{D_i(t-,\mat)}
    \la_{i,j} (t)
    +
    \la_{i,i} (t)
    \bigg)dt
    \\ \label{eq:JA6}
    & =
    \sum_{i=1}^{K-1}
    H_i(t)
    \f{D_i(t-,\mat)}{B_t}
    \bigg(
    \sum_{j=1, j \neq i}^{K-1}
    \bigg[
    \f{D_j(t-,\mat)}{D_i(t-,\mat)} -1
    \bigg]
    \la_{i,j} (t)
    -
    \la_{i,K} (t)
    \bigg)dt .
\end{align}
 Since $H_i(t)=1$ on the set $\{ C^1(t) = i\}$ and zero on its
complement we can write
\begin{align*} \nonumber
    I_1& + I_2 =  (1 - H_K(t))
    \f{D_{C^1(t)}(t-,\mat)}{B_t}
    \bigg(
    g_{C^1(t)}(t,t) - f(t,t)
    -
    (1- \d_{C^1(t)})
    \la_{C^1(t),K} (t)
     + {\bar{a}_{C^1(t)}}(t, \mat)
     \\
    & +
    \d_{C^1(t)}
    \bigg[
    \f{B(t-, \mat)}{D_{C^1(t)}(t-, \mat)}
    -1
    \bigg]
    \la_{C^1(t),K}(t)
     +
    \sum_{j= 1,j \neq C^1(t)}^{K-1}
    \bigg[
    \f{D_j(t-,\mat)}{D_{C^1(t)}(t-,\mat)} -1
    \bigg]
    \la_{C^1(t),j} (t)
    \bigg) dt ,
 \end{align*}
 where we have also used the fact that we sum only up to
$K-1$. We conclude the argument as in Theorem
\ref{thm:frac-market-mig}. \kd

{\bf Proof of Theorem \ref{thm:par-mig}. \ }
 We have
 \beqa
        \label{eq:HJM1credit_mig-nJ}
    d  \Big( \f{D(t,\mat)}{B_t} \Big)
    =
    \sum_{i=1}^{K-1}\ \Big(  d \Big( H_i(t) \f{D_i(t,\mat)}{B_t} \Big) + d \Big( H_{i,K}(t)
      \f{\d_i}{B_\tau} \Big) \Big) .
 \eeqa
The first part was calculated before (see  \eqref{eq:JA3}). The
second part can be written using the martingale $M_{i,K}$ as
\[
    \f{\d_i}{B_t} d H_{i,K}(t) = \f{\d_i}{B_t} d M_{i,K}(t) + \f{\d_i}{B_t} H_i(t) \la_{i,K}(t)
    dt.
\]
Hence by  \eqref{eq:JA6}  the drift term $I$ of
\eqref{eq:HJM1credit_mig-nJ} is given by

\begin{align*}
   I = & \sum_{i=1}^{K-1} \
   \bigg(
    \f{D_i(t-,\mat)}{B_t}
    \Big(
    \la_{C^1(t),i} (t)
    +
    H_i(t-)(g_i(t,t) - f(t,t) + {\bar{a}_i}(t, \mat))
    \Big)dt
    +
    \f{\d_i}{B_t} H_i(t) \la_{i,K}(t) dt \ \bigg)  \\
&  = \sum_{i=1}^{K-1}
    H_i(t-)
    \f{D_i(t-,\mat)}{B_t}
    \bigg(
    g_i(t,t) - f(t,t) + {\bar{a}_i}(t, \mat)
    +
    \d_i\bigg[
    \f{1}{D_i(t-,\mat)} -1
    \bigg]\la_{i,K}(t)   \\
    & +
        \sum_{j=1,j \neq i}^{K-1}
    \bigg[
    \f{D_j(t-,\mat)}{D_i(t-,\mat)} -1
    \bigg]
    \la_{i,j} (t)
    -
    (1 - \d_i)
    \la_{i,K} (t)
    \bigg)dt \\
    &  = (1- H_K(t))
    \f{D_{C^1(t)}(t-,\mat)}{B_t}
    \bigg(
    g_{C^1(t)}(t,t) - f(t,t) +  {\bar{a}_{C^1(t)}}(t, \mat) -
    (1 - \d_{C^1(t)})
    \la_{C^1(t),K} (t) \\
& +
    \d_{C^1(t)}\bigg[
    \f{1}{D_{C^1(t)}(t-,\mat)} -1
    \bigg]\la_{C^1(t),K}(t)
    +
    \sum_{j=1, j \neq C^1(t)}^{K-1}
    \bigg[
    \f{D_j(t-,\mat)}{D_{C^1(t)}(t-,\mat)} -1
    \bigg]
    \la_{C^1(t),j} (t)
    \bigg)dt .
 \end{align*}
 Arguing as before we complete the proof.
\kd

{\bf Proof of Theorem \ref{th:multiple}. \ } By the It\^o lemma
  \begin{align}  \label{eq:po26}
     d \bigg( \f{D(t,\mat)}{B_t}\bigg)
     = &
     V_{t-} \sum_{i=1}^{K-1} d \bigg( H_i(t)\f{D_{i} (t,\mat)}{B_t}
     \bigg)
      +
     \bigg( \sum_{i=1}^{K-1} H_i(t)\f{D_i (t,\mat)}{B_t}
     \bigg)
     d V_t
     \\  \nonumber & + \De \bigg( \sum_{i=1}^{K-1} H_i(t) \f{D_{i} (t,\mat)}{B_t} \bigg) \De
     V_t  = I_1 + I_2  + I_3.
 \end{align}
 By assumption (H2) we have $I_3=0$. \\
 By  \eqref{eq:schon1},  \eqref{eq:schon2} and the fact that
$D_i(\cdot, \mat)$ is a \cadlag process, we obtain
  \begin{align*}
    - I_2 &=
         \bigg(
     \sum_{i=1}^{K-1} H_i(t)\f{D_{i} (t,\mat)}{B_t}
     \bigg)
      V_{t-} L_t d M_t
      +
     \bigg(
     \sum_{i=1}^{K-1} H_i(t)\f{D_{i} (t,\mat)}{B_t}
     \bigg)
      V_{t-} L_t  \gamma_t   d t
     \\
      & =
     \bigg(
     \sum_{i=1}^{K-1} H_i(t)\f{D_{i} (t,\mat)}{B_t}
     \bigg)
      V_{t-} L_t d M_t
      +
     \bigg(
     \sum_{i=1}^{K-1} H_i(t)\f{D_{i} (t-,\mat)}{B_t}
     \bigg)
      V_{t-} L_t   \gamma_t d t .
\end{align*}
Hence, taking into account (\ref{eq:po26}),  (\ref{eq:JA3}), we
see that the drift term of $ d(\f{D(t,\mat)}{B_t})$ is given by
 $$    \sum_{j=1}^{K-1}
     V_{t-} \f{D_j(t-,\mat)}{B_t}
     \la_{C^1(t),j} (t) dt
     +
     \sum_{i=1}^{K-1}
         H_i(t)
         V_{t-} \f{D_i(t-,\mat)}{B_t}
     \bigg(
       g_i(t,t) - f(t,t) + {\bar{a}_i}(t, \mat)
       - L_t   \gamma_t
     \bigg)
     dt .
$$ Since
  \begin{align*}
     & \sum_{j=1}^{K-1}
     V_{t-} \f{D_j(t-,\mat)}{B_t}
     \la_{C^1(t),j} (t) =
     \sum_{i=1}^{K-1}
     H_i(t)
     V_{t-} \f{D_i(t-,\mat)}{B_t}
     \sum_{j=1}^{K-1}
     \f{D_j(t-,\mat)}{D_i(t-,\mat)}
     \la_{i,j} (t)  \\
     & =
     \sum_{i=1}^{K-1}
     H_i(t)
     V_{t-} \f{D_i(t-,\mat)}{B_t}
     \bigg(
     \sum_{j=1, j \neq i}^{K-1}
     \f{D_j(t-,\mat)}{D_i(t-,\mat)}
     \la_{i,j} (t)
     +
     \la_{i,i} (t)
     \bigg)
     \\
     &=
     \sum_{i=1}^{K-1}
     H_i(t)
     V_{t-} \f{D_i(t-,\mat)}{B_t}
     \sum_{j=1, j \neq i}^{K-1}
     \bigg[
     \f{D_j(t-,\mat)}{D_i(t-,\mat)}
     -1
     \bigg]
     \la_{i,j} (t) ,
   \end{align*}
the drift term is given by
  \begin{align*}
     \sum_{i=1}^{K-1}
         H_i(t)
         V_{t-} \f{D_i(t-,\mat)}{B_t}
     \bigg(
            g_i(t,t) - f(t,t) + {\bar{a}_i}(t, \mat)
       - L_t  \gamma_t
      +
     \sum_{j=1, j \neq i}^{K-1}
     \bigg[
     \f{D_j(t-,\mat)}{D_i(t-,\mat)}
     -1
     \bigg]
     \la_{i,j} (t)
     \bigg)
     dt.
   \end{align*}
  Arguing as in the previous sections, we obtain the theorem.
\kd

{\bf Proof of Theorem \ref{thm:D-HJM-type-cond}. \ } Theorem
\ref{thm:D-HJM-type-cond}
follows from Lemma  \ref{lem1} by using the following facts on derivatives (the details are left to the reader): \\
 i)  for fractional recovery of Treasury value  \beq
        \f{\prt}{\prt \mat}\bigg( \f{B(t-,\mat)}{D_1(t-,\mat)}   -1  \bigg)
        &=&
        \Big(
        g_1(t-,\mat)
        -f(t-,\mat)
        \Big)
        e^{ \int_t^\mat (g_1(t-,u) - f(t-,u)) du},
    \eeq
ii) for fractional recovery of par value
    \beq
        \f{\prt}{\prt \mat} \bigg(  \f{1}{D_1(t- , \mat)} -1 \bigg)
        &=& g_1(t-,\mat)
        e^{ \int_t^\mat g_1(t-,u) du }  ,
    \eeq
 iii)  for fractional recovery of market value with rating migration
    \beq
        \f{\prt}{\prt \mat}
       \bigg(
       \f{D_i(t-,\mat)}{D_{C^1(t)}(t-,\mat)}
       -1
       \bigg)
       =
       \Big(
       g_{C^1(t)}(t-,\mat)
       -g_i(t-,\mat)
       \Big)
       e^{ \int_t^\mat ( g_{C^1(t)}(t-,u) - g_i(t-,u) ) du },
    \eeq
 iv)   for fractional recovery of treasury value with rating migration
    \beq
        \f{\prt}{\prt \mat} \bigg( \f{B(t-,\mat)}{D_{C^1(t)}(t-,\mat)}   -1  \bigg)
        &=&
        \Big(
        g_{C^1(t)}(t-,\mat)
        -f(t-,\mat)
        \Big)
        e^{\int_t^\mat (g_{C^1(t)}(t-,u) - f(t-,u)) du },
    \eeq
 v) for fractional recovery of par value with rating migration
    \beq
        \f{\prt}{\prt \mat} \bigg(  \f{1}{D_{C^1(t)}(t- , \mat)} -1 \bigg)
        &=&
        g_{C^1(t)}(t-,\mat)
        e^{\int_t^\mat g_{C^1(t)}(t-,u) du } .
    \eeq
\kd
{\bf Proof of Theorem \ref{thm:cons-HJM-type-cond}. \ }
Under hypothesis (H1) we can write  \eqref{eq:cons-market-value} in the form
  \beqa
    \nonumber
    && \sum_{i=1, i \neq C^1(t)}^{K-1}
    \bigg[
    \big(
    D_i(t-,\mat)
    -D_{C^1(t)}(t-,\mat)
    \big)
    \la_{C^1(t),i} (t)  \bigg] + \\
    && \nonumber -
    ( 1- \d_{C^1(t)}(t)) D_{C^1(t)}(t-,\mat)
    \la_{C^1(t),K}(t) \\
    && \nonumber
        +
    \left(( 1 - \delta_{ C^1(t) }(t) ) \lambda_{C^1(t),K} (t) +
    \bar{a}_{C^1(t)}(t,\mat) \right)D_{C^1(t)}(t-, \mat)    =0
.
\eeqa
By definition of $\bar{a}_i(t,\mat)$ we see that this condition is equivalent to
  \beqa
    \sum_{i=1, i \neq C^1(t)}^{K-1}
    \left(
\frac{
    D_i(t-,\mat)}{ D_{C^1(t)}(t-,\mat) }
    - 1
    \right)
    \la_{C^1(t),i} (t)
         -
        \int_t^\mat
        \a_{C^1(t)}(t,u) du
        + J_{C^1(t)} \left(\int_t^\mat \s_i(t,v) dv \right)
 =0
\eeqa
which is exactly the condition \eqref{eq:HJM2credit_mig}.

\bigskip

\n {\bf Acknowledgements}. The authors  wish to thank  Jerzy
Zabczyk for stimulating discussions and comments during the
writing of this paper.

\bibliographystyle{plain}

\end{document}